\documentclass[aps,pra,twocolumn,showpacs,groupedaddress,floatfix]{revtex4}

\usepackage[english]{babel}

\usepackage{amsmath}
\usepackage{bm}

\usepackage{graphicx}

\usepackage{qphys}
\newcommand{\e}{\ensuremath{\mathrm{e}}}

\begin{document}

\title{Growth of graph states in quantum networks}
\author{Mart\'\i\ Cuquet}
\author{John Calsamiglia}
\affiliation{F\'{\i}sica Te\`orica: Informaci\'o i Fen\`omens
Qu\`antics, Departament de F\'{\i}sica, Universitat Aut\`onoma de Barcelona, 08193 Bellaterra, Barcelona,
Spain}
\date{\today}

\begin{abstract}
  We propose a scheme to distribute graph states over quantum networks in the presence of noise in the channels and in the operations.
  The protocol can be implemented efficiently for large graph sates of arbitrary (complex) topology.
  We benchmark our scheme with two protocols where each connected component is prepared in a node belonging to the component and subsequently distributed via quantum repeaters to the remaining connected nodes.
  We show that the fidelity of the generated graphs can be written as the partition function of a classical Ising-type Hamiltonian.
  We give exact expressions of the fidelity of the linear cluster and results for its decay rate in random graphs with arbitrary (uncorrelated) degree distributions.
\end{abstract}


\maketitle

\section{Introduction}

Quantum networks \cite{cirac_quantum_1997,Kimble2008}, where nodes with (limited) quantum storage and processing power are coupled through quantum channels, are becoming a focus of interest in quantum information.
The first motivation is to extend the paradigmatic bipartite quantum communication applications, e.g., quantum teleportation \cite{bennett_teleporting_1993} or more prominently quantum key distribution \cite{gisin_quantum_2002}, to a multipartite setting, where such bipartite protocols can be accomplished between arbitrary nodes of the network.
The possibility of networks with non-trivial topologies can give rise to new phenomena and to applications that exploit the multipartite correlations.
Of course, this puts forward a big variety of theoretical and technological challenges which can be addressed in short term.
Indeed the first steps have already been taken towards a quantum network in a first realization with two distant nodes that can store and interchange quantum information in an efficient and reversible way \cite{ritter_elementary_2012}.
Lastly, such physical realizations are in principle scalable and hence open the door to perform highly controllable experiments on many-body phenomena, study multipartite entanglement \cite{jungnitsch_taming_2011}, and could eventually perform more complex tasks like some implementation of the quantum Google page rank \cite{paparo_google_2012} or general distributed quantum computations.

A central task in quantum networks is to devise strategies to distribute entanglement among its nodes.
Linear networks have shown to be useful for long-distance bipartite entanglement distribution by means of quantum repeaters \cite{Briegel1998} (see also its measurement-based implementation  \cite{zwerger_measurement-based_2012}).
However, the study of entanglement distribution over higher dimensional networks is in its infancy.
Current results show that in some scenarios the network topology can bring interesting effects like entanglement percolation \cite{acin_entanglement_2007,cuquet_entanglement_2009,perseguers_fidelity_2010,perseguers_quantum_2010, broadfoot_singlet_2010,cuquet_limited-path-length_2011,broadfoot_entanglement_2009} that lead to new approaches to the problem.
This effect shows up in non-ideal (but still not fully realistic) scenarios or for some particular network geometries.

In this paper we address the problem of distributing graph states in the realistic scenario of noisy network channels and a small, but non-negligible, amount of noise in the local operations.
Graph states \cite{Hein2004} are a large family of multipartite entangled states that, although they can be efficiently described with relatively few parameters, have a rich variety of features.
For instance, they include paradigmatic states like GHZ, cluster states, codewords of error-correcting codes \cite{schlingemann_quantum_2001}, they provide some novel quantum communication applications, like secret entangled-state distribution \cite{dur_multipartite_2005}, they are useful in the study of non-locality \cite{guhne_bell_2005}, and most importantly they include states which are universal resources for measurement-based quantum computing \cite{briegel_measurement-based_2009,van_den_nest_universal_2006}.

Our goal here is to propose a protocol to create a large graph using an underlying network of noisy channels.
The protocol should tolerate channels and operations with errors, scale efficiently with the size of the network, and work for any network topology, and in particular for complex networks.
Complex networks~\cite{Albert2002,Dorogovtsev2008,Newman2010} underlie many natural \cite{Jeong2001,Scala2001}, social \cite{Watts1998,Amaral2000} and artificial systems \cite{Pastor-Satorras2001_dynamical,Yook2002} where different parties interact.
Their non-trivial structure is the source of features, like the existence and shape of an important fraction of highly connected nodes \cite{Newman2001_random} and the tendency of nodes to cluster together \cite{Watts1998}, that are often observed in real graphs and have a deep impact in their performance.
Complex networks are of particular importance in communication infrastructures, as most present telecommunication networks like Internet have a complex structure \cite{Pastor-Satorras2001_dynamical,Yook2002}.
However, the rich and intriguing properties of these type of networks are still quite unexplored in the quantum setting (see nonetheless \cite{cuquet_entanglement_2009,perseguers_quantum_2010,cuquet_limited-path-length_2011,wiersma_random_2010, mulken_continuous-time_2011}).

Channels linking separate nodes in a network are typically noisy, and pose the main caveat to the creation of distributed multipartite entangled states with high fidelity.
To overcome this, there exist bipartite \cite{Bennett1996,Deutsch1996} and multipartite \cite{Dur2003,Aschauer2005,Kruszynska2006} entanglement purification protocols that allow either to generate highly purified Bell pairs, which can latter be used to teleport an arbitrary graph state, or to directly purify the desired graph state.
These recursive protocols tolerate a reasonable amount of noise in local operations, but require a number of initial copies that grows exponentially with the size of the state.
Other proposals do not use postselection, making the purification efficient in terms of the size of the graph state, but come at the expense of a stricter noise threshold \cite{Goyal2006}.
In Ref.~\cite{Campbell2007}, an entanglement pumping without postselection is also used to obtain efficient purification when constructing the graph state edge by edge.
In different approaches, the graph state is created by a probabilistic growth using non-deterministic entangling operations \cite{Kieling2007a,Rohde2007,Campbell2007b,Campbell2007a,Matsuzaki2010}.

Here we investigate the advantages of generating and purifying small GHZs that reproduce the local structure of the network, and merge them in order to distribute a network-wide graph state.
Since GHZ have a fixed size which depends on the degree of each node and is thus independent of the size of the network, this protocol is efficient in the size while it still maintains the high thresholds of the recurrence schemes.
We benchmark this protocol with two other protocols that generate high-fidelity bipartite states between a node and the rest of the network, which are then used to distribute a locally generated graph state.
We use the fidelity of the graph state as a figure of merit to compare the three protocols.
The fidelity decays exponentially with the size of the network for a constant level of noise, and hence we also use its decay rate.
It turns out that both quantities can be understood as the partition function and free energy of a thermodynamic system, respectively, and thus standard methods of statistical mechanics are readily used.

The paper is structured as follows.
First, in Section~\ref{sec:def} we introduce graph states and the noise model we consider.
In Section~\ref{sec:protocols} we present the three protocols and the two figures of merit---the fidelity of the graph state and its decay rate---, and relate them with an analog partition function and free energy of a thermodynamical system.
Then, in Section~\ref{sec:nets} we apply the protocols to the creation of a linear cluster state, for which we obtain exact results, and of a graph state associated to a complex network.
We conclude in Section~\ref{sec:conc}.

\section{\label{sec:def}Definitions}

\subsection{Graph states}

A graph $G=\{V(G),E(G)\}$ is an ordered pair of sets: the set $V(G)$ of vertices, or nodes, and the set $E(G)$ of edges, or links, whose elements are (unordered) pairs of vertices and represent the connections between them.
The neighborhood of a vertex $u\in V(G)$ is the set of vertices connected to it, $\mathcal{N}_u(G) = \{v:(u,v)\in E(G)\}$, and the degree of that vertex is the number of its neighbors, $k=|\mathcal{N}_u(G)|$.
If the graph is \emph{directed}, then the elements of $E(G)$ are ordered, and $(u,v)\in E(G)$ is a directed edge from $u$ to $v$.
In this case, the incoming and outgoing neighborhoods of $u$ are $\mathcal{N}^{\rm (in)}_u(G) = \{v:(v,u)\in E(G)\}$ and $\mathcal{N}^{\rm (out)}_u(G) = \{v:(u,v)\in E(G)\}$, respectively.

A graph state \cite{Hein2004} is a quantum state associated to a graph $G$, where vertices correspond to qubits and edges to interactions.
A common description of graph states is the stabilizer formalism.
For a graph $G$, there are $N=|V(G)|$ stabilizer operators (one for each vertex $u$) defined as
\begin{equation}
  K_u^G = X_{u} \prod_{v\in\mathcal{N}_u(G)} Z_{v}
  .
  \label{eq:stabilizer}
\end{equation}
Here $X$, $Y$ and $Z$ are the Pauli operators and the subindex denotes on which qubit they operate.
A pure graph state $\ket{\bm{\mu}}_G$, with $\bm{\mu}=\{0,1\}^N$, is a common eigenstate of all stabilizer operators with
\[
  K_u^G \ket{\bm{\mu}}_G = (-1)^{\mu_u} \ket{\bm{\mu}}_G \quad \forall u\in V
  ,
\]
where $\mu_u$ is the $u$-th component of $\bm{\mu}$.
The set $\{\ket{\bm{\mu}}_G\}$ form the graph state basis.
We will omit $G$ when the graph is clear by context.

An alternative description is by means of the interaction picture, where $\ket{\bm{0}}_G$ is created by preparing all qubits in the state $\ket{+}$ and then applying a $\textsc{cphase}_{u,v}$ for every edge $(u,v)\in E$,
\[
  \ket{\bm{0}}_G = \prod_{(u,v)\in E} \textsc{cphase}_{u,v} \ket{+}^{\otimes N}.
\]
Since $Z_{u}$ anticommutes with $K_u$ and commutes with the rest of $K_v$, $v\neq u$, any graph-basis element can be expressed as
\[
  \ket{\bm{\mu}}_G = \prod_{u} \left( Z_{u} \right)^{\mu_u} \ket{\bm{0}}_G.
\]

Throughout this paper we consider mixed graph states diagonal in the graph state basis, $\rho_G = \sum_{\bm{\mu}} \lambda_{\bm{\mu}} \ketbra{\bm{\mu}}{\bm{\mu}}$.
Any mixed graph state can be brought to this form by local depolarization \cite{Aschauer2005}.
We will work however in the stabilizer basis.
Let us define $\bm{K}_{\bm{x}}=\prod_{u\in V} \left( K_v \right)^{x_v}$, where $\bm{x} = (x_1x_2\dots x_N)\in\{0,1\}^N$.
Since $\{\bm{K}_{\bm{x}}\}$ form a complete set of commuting observables, a diagonal graph state can be expressed as $\rho_G =\frac{1}{2^{N}} \sum_{\bm{x}} \left\langle \bm{K}_{\bm{x}} \right\rangle \bm{K}_{\bm{x}}$, where $\expected{\bm{K}_{\bm{x}}} = \sum_{\bm{\mu}}\lambda_{\bm{\mu}}(-1)^{\bm{\mu}\cdot\bm{x}}$.
The effect of the Pauli operators on $\bm{K}_{\bm{x}}$ is
\begin{align}
  Z_{u} \bm{K}_{\bm{x}} Z_{u} &= (-1)^{x_u} \bm{K}_{\bm{x}}, \\
  X_{u} \bm{K}_{\bm{x}} X_{u} &= \prod_{v\in\mathcal{N}_u}(-1)^{x_v} \bm{K}_{\bm{x}}, \\
  Y_{u} \bm{K}_{\bm{x}} Y_{u} &= (-1)^{x_u} \prod_{v\in\mathcal{N}_u}(-1)^{x_v} \bm{K}_{\bm{x}},
\end{align}
so this unitaries map diagonal graph states into diagonal graph states.
A $\textsc{cphase}_{u,v}$ adds an edge between $u$ and $v$, if there where not connected, or removes it, if the edge already existed.
The effect on $\bm{K}_{\bm{x}}$ is
\begin{equation}
  \textsc{cphase}_{u,v} \bm{K}_{\bm{x}} \textsc{cphase}_{u,v}
  =
  \bm{K}_{\bm{x}} \left( Z_v \right)^{x_u} \left( Z_u \right)^{x_v}
  .
\end{equation}

The action of Pauli measurements can also be easily described in this formalism as a transformation of the graph (up to some local unitaries).
Measurement of $Z$ simply disconnects the measured qubit from the rest of the graph, while $X$ and $Y$ transform the neighborhood of the measured qubit and then disconnect it.
In terms of the stabilizer operators, the measurement of $Z_u$ commutes with all $K_v$, for $v\neq u$, and anticommutes with $K_u$.
Thus,
\begin{multline}
  \left[ I + (-1)^m Z_u \right] \bm{K}_{\bm{x}} \left[ I + (-1)^m Z_u \right]
  \\
  =
  \left[ I + (-1)^m Z_u \right] \bm{K}_{\bm{x}} \delta_{0,x_u}
  ,
\end{multline}
where $m=\{0,1\}$ labels the measurement outcome $\{+1,-1\}$ respectively.
After tracing out qubit $u$, the new stabilizer is
\begin{equation}
  (-1)^{m\cdot\sum_{v\in\mathcal{N}_u}x_v} \bm{K}\pr_{\bm{x}} \delta_{0,x_u},
  \label{sZ}
\end{equation}
where the new $\bm{K}\pr_{\bm{x}}$ corresponds to a new graph $G\pr$ obtained form $G$ by removing vertex $u$ and its attached edges.
The spurious phase factor in \eqref{sZ} can be cancelled by applying a unitary $\left( \prod_{v\in\mathcal{N}_u} Z_v \right)^m$.
Similarly, measurements of $X_u$ or $Y_u$ also result in the disconnection of the measured qubit, but in these cases the remaining graph is transformed by local complementations of the neighborhood of $u$, as described in \cite{Hein2004}.
We will be more explicit in the concrete cases where we apply these measurements.

Our figure of merit will be given in terms of the fidelity, which for a graph-diagonal state can be written as,
\begin{equation}
  F^G_N
  =
  {_G}\!\bra{\bm{0}}\rho_{G} \ket{\bm{0}}_G
  =
  \lambda_{\bm{0}}
  =
  \frac{1}{2^N}\sum_{\bm{x}}\expected{\bm{K}^G_{\bm{x}}}
  .
\end{equation}

\subsection{Network and noise model}

We consider a network where nodes are spatially separated.
Some of the nodes are connected by a link, which we model as a noisy depolarizing channel on one qubit with error parameter $p_c$,
\begin{equation}
  T_c^{(u)} = (1-p_c) [I] + \frac{p_c}{4} \sum_{i=0}^3 [\sigma_i^{(u)}]
  .
  \label{eq:noise_channel}
\end{equation}
Here $\sigma_i$ are $I$, $X$, $Y$, $Z$ for $i=0,1,2,3$, respectively.
A Pauli measurement on $u$ is modeled as a perfect measurement preceded by a depolarizing channel $T_1^{(u)}$ with error probability $p_1$ on that qubit,
\begin{equation}
  T_1^{(u)} = (1-p_1) [I] + \frac{p_1}{4} \sum_{i=0}^3 [\sigma_i^{(u)}]
  .
  \label{eq:noise_pauli}
\end{equation}
A noisy two-qubit gate (e.g., a \textsc{cphase} or a \textsc{cnot}) on qubits $u$ and $v$ is modeled as an ideal gate followed by the two-qubit depolarizing channel on $u,v$ with error parameter $p_2$,
\begin{equation}
  T_2^{(u,v)} = (1-p_2) [I] + \frac{p_2}{16} \sum_{i,j=0}^3 [\sigma_i^{(u)}\otimes\sigma_j^{(v)}]
  \label{eq:noise_2qubit}
\end{equation}
Gates can only be applied locally, i.e., on qubits within the same node.

In the stabilizer basis, the effect of each noise source is easily tracked: it multiplies each stabilizer element by a coefficient, and keeps the graph state in diagonal form.
A noise $T_2^{(u,v)}$ changes the sign of the coefficient accompanying $\bm{K}_{\bm{x}}$ with probability $p_2/2$ if it acts nontrivially on $u$ or $v$, i.e., if exists at least one $x_a=1$ for $a\in u\cup v\cup\mathcal{N}_u\cup\mathcal{N}_v$.
Thus, each $T_2^{(u,v)}$ adds a coefficient
\begin{equation}
  (1-p_2)^{\theta(x_u,x_v,\bm{x}_{\mathcal{N}_u},\bm{x}_{\mathcal{N}_v})}
  ,
  \label{eq:noise_2qubit_coeff}
\end{equation}
where $\theta(\bm{x})\equiv1-\delta(\bm{0},\bm{x})$ and $\bm{x}_{\mathcal{N}_u}=(x_{v_1}\dots x_{v_k})$ for all $v_a\in\mathcal{N}_u$.
Noise $T_1^{(u)}$ behaves similarly.
In this case, the sign is changed with probability $p_1/2$ unless $x_u=0$ and $\bigoplus_ax_a=0$ for $a\in\mathcal{N}_u$, so the multiplying coefficient is
\begin{equation}
  (1-p_1)^{\theta(x_u,\bigoplus_{v\in\mathcal{N}_u}x_v)}
  .
  \label{eq:noise_pauli_coeff}
\end{equation}
Note thus that these noise sources affect the qubits on which the gates act, plus their neighbors.

Finally let us point that we do not associate any noise to the local ``correcting'' unitaries performed in order to bring the post-measurement states to a standard graph form.
We refer to this unitaries in our protocols to simplify bookkeeping, but their action can be pushed forward (or commuted) till the end of the protocol, and hence the resulting state is exactly equivalent as a resource of entanglement.

\section{\label{sec:protocols}Protocols}

We propose a protocol to distribute a graph state with the structure of a general quantum communication network, of arbitrary topology, associated to graph $G$.
Noise $p_c$ in the communication channels is considered to be relatively high, so some sort of purification or error correction is in order.
The protocol uses the structure of the network to distribute several copies of small subgraphs between neighbors, and purifies them by means of multipartite purification.
These subgraphs are GHZ states, which are associated to a star graph with a central node of degree $j$, connected to $j$ leaves of degree $1$.
In order to benchmark our protocol, we also consider two reference protocols that distribute bipartite states between a central node, which locally creates the desired graph state, and the rest of the network.
The bipartite states are then used to teleport the locally created graph state.
Since this central node may not be directly connected to the rest of the network, quantum repeaters \cite{Briegel1998} are used to establish purified bipartite states between this node and all the network's nodes.

In all cases, we consider the same multipartite purification protocol for bicolorable graph states described in Ref.~\cite{Aschauer2005}.
The noise threshold of this purification scheme depends on the degree of the central qubit of the state that is purified, and goes from $\approx0.06$ for a bipartite state (i.e., a GHZ with central degree $1$) to $\approx0.02$ for a GHZ of central degree $9$.
This poses a limit in the maximum degree of a network for which the protocols can be used.
Hence, we work under the assumption that the errors in the local operations are low---as compared to the (finite) degree of the subgraphs, $j p\ll 1$---and that the channel noise can be significant, but also low enough so that the minimum threshold fidelity required for the purification protocol to succeed can be attained.
At first order in $p_1=p_2=p$, the output state fixed point of this purification scheme is given by (see Appendix~\ref{sec:multipartite_protocol})
\begin{equation}
 \expected{\!K_a^{x_a}\!\!\prod_{b\in\mathcal{N}_a} K_b^{x_b}\! \!}
  =
  1\!-\!p \left[ \overline{x_a} \left\lceil \frac{\sum_b x_b}{2} \right\rceil\! +\! x_a (j+1)\right]
  ,
  \label{eq:multipartite_fixed}
\end{equation}
where $a$ is the central node and $b$ the leaves of the GHZ, $j$~is the number of leaves and the overline in $x_a$ represents the bit-complement, $\overline{x_a}=x_a\oplus1$.
 
Note that the extensive use of quantum repeaters in the first two protocols renders them extremely inefficient.
Nevertheless, we will find that the performance in terms of attainable fidelity is still comparable to that of our more efficient subgraph protocol.

\subsection{Bipartite A protocol}

\begin{figure}[tb]
  \begin{center}
    \includegraphics[width=.45\textwidth]{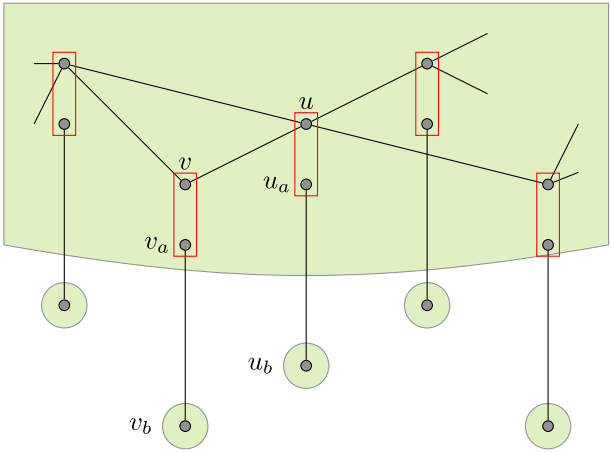}
  \end{center}
  \caption{Bipartite A protocol.
  The upper green area represents the ``central'' node, where the local graph state with the structure of the network is generated.
  The lower green circles are nodes in the network.
  Small, gray dots correspond to qubits, and lines connect neighbors.
  Rectangles in red indicate Bell measurements involved in the teleportation of the local graph states, which are implemented as \textsc{cphase}s between $u$ and $u_a$ and $X$ measurements at $u$, $u_a$.}
  \label{fig:channels1}
\end{figure}
In the first protocol, purified, but still noisy, entangled states are created between a central node and all the others by means of quantum repeaters.
This central node then teleports a locally generated state to all the other nodes.
For simplicity, in protocols $A$ and $B$ we assume that the central node is sending \emph{all} qubits through purified channels, including his own.
This adds a source of noise that would not be strictly necessary, but its effect is small for large enough networks.
In cases, like complex networks, where the network is not necessarily connected, it is understood that there is a central node for each connected component distributing the corresponding graph state.

The local state $\rho_G$, which mimics the structure of the network, is created by initializing $N$ qubits in the state stabilized by $X$ and then applying \textsc{cphase}s between neighbors.
Then, the noisy Bell states between the central node and all the others are used to teleport the corresponding qubits (see Fig.~\ref{fig:channels1}).
We label qubits in the local graph as $u\in V(G)$.
For each node in the network, there is a bipartite state $\rho_{g_u}$ of two qubits, $u_a,u_b\in V(g_u)$ and one edge $(u_a,u_b)\in E(g_u)$.
Qubit $u_a$ belongs to the central node, while $u_b$ is in the corresponding node in the network.
In order to account for the errors in the teleportation Bell--measurement, we implement it by a \textsc{cphase} on $(u,u_a)$ followed by $X$ measurements on $u$ and $u_a$.

Without taking noise into account, the state before teleportation is
\begin{equation}
\label{eq:rhobeforet}
  \rho_G \otimes \bigotimes_{u\in V(G)} \rho_{g_u}
\end{equation}
with
\begin{equation}
\label{eq:rhobeforet2}
\rho_G
  =
  \frac{1}{2^{N}}\sum_{\bm{x}} \bm{K}_{\bm{x}}^G  \mbox{ and }
  \rho_{g_u}
  =
  \frac{1}{4} \sum_{x_{u_a},x_{u_b}}
  (K_{u_a}^{g_u})^{x_{u_a}} (K_{u_b}^{g_u})^{x_{u_b}}
  .	\nonumber
\end{equation}
The graphs $g_u$ correspond to the bipartite states with qubits $u_a$ and $u_b$ used for teleportation, i.e., $K_{u_a}^{g_u}=X_{u_{a}} Z_{u_{b}}$.
The action of the $\textsc{cphase}_{u,u_a}$ affects only the stabilizers $K_u^G\to K_u^GZ_{u_a}$ and $K_{u_a}^{g_u}\to K_{u_a}^{g_u}Z_{u}$.
The measurement of $X_{u}$ anticommutes with all $K^G_v$, $v\in\mathcal{N}_u(G)$, and with $K_{u_a}^{g_u}Z_{u}$, while that of $X_{u_a}$ anticommutes only with $K_u^GZ_{u_a}$ and $K_{u_b}^{g_u}$.
Thus, each term changes to
\begin{multline}
  \left[ 1+(-1)^{m_{u_a}}X_{u_a} \right]
  \left[ 1+(-1)^{m_u}X_{u} \right]
  \bm{K}_{\bm{x}}
  \\
  \left( Z_{u_a} \right)^{x_u}
  \left( K_{u_a}^{g_u}Z_{u} \right)^{x_{u_a}}
  \left( K_{u_b}^{g_u} \right)^{x_{u_b}}
  \\
  \delta_{0,x_{u_a}\oplus\bigoplus_{v\in\mathcal{N}_u}x_v}
  \delta_{0,x_{u}\oplus x_{u_b}}
  ,
\end{multline}
where $m_u$ and $m_{u_a}$ are the measurement outcomes.
Tracing out qubits $u$ and $u_a$ and correcting the state with
$\left( Z_{u_b} \right)^{m_u} \left( X_{u_b} \right)^{m_{u_a}}$
we obtain
$
  \bm{K}_{\bm{x}}\pr \delta_{0,x_{u_a}\oplus\bigoplus_{v\in\mathcal{N}_u}x_v}
  \delta_{0,x_{u}\oplus x_{u_b}}
$,
with $\bm{K}_{\bm{x}}\pr$ associated to a new graph $G\pr$ where vertex $u$ has been substituted by vertex $u_b$.
Teleportation of all local qubits results in the desired distributed state.

Noise can now be introduced as the multiplicative factors in front of the stabilizer elements of \eqref{eq:rhobeforet}.
Here, the order of the \textsc{cphase} gates used to generate the local graph matters, as qubits receive noise from gates performed at neighbors which are already connected to them.
There is thus a noise $(1-p_1)^{x_u}$ corresponding to the preparation of each node in $X_u$, and a $(1-p_2)^{\theta(x_u,x_v,x_{\tilde{\mathcal{N}}_u},x_{\tilde{\mathcal{N}}_v})}$ for each edge in the local state, where the tilde in $\tilde{\mathcal{N}}_u$ labels that we consider the neighborhood of $u$ at the moment the $\textsc{cphase}_{u,v}$ is performed.
Additionally, and using that $x_{u_a}=\bigoplus_{v\in\mathcal{N}_u}x_v$ and $x_{u_b}=x_u$, in each teleportation the \textsc{cphase} introduces noise as $(1-p_2)^{\theta(x_u,\bm{x}_{\mathcal{N}_u})}$, and the measurements as $(1-p_1)^{x_u+\bigoplus_{v\in\mathcal{N}_u}x_v}$.
Finally, there is a $\expected{(K_{u_a}^{g_u})^{\bigoplus_{v\in\mathcal{N}_u}x_v}(K_{u_b}^{g_u})^{x_u}}$ term from the purified Bell states.
This results in a final distributed state $\rho = \frac{1}{2^N}\sum_{\bm{x}} \expected{\bm{K}_{\bm{x}}} \bm{K}_{\bm{x}}$ with
\begin{align}
  \expected{\bm{K}_{\bm{x}}}
  &=
  \prod_{u\in V}
  \left[
  (1-p_1)^{x_u}
  \expected{(K_{u_a}^{g_u})^{\bigoplus_{v\in\mathcal{N}_u}x_v}(K_{u_b}^{g_u})^{x_u}}
  \right.
  \notag\\
  &\qquad
  \left.
  (1-p_2)^{\theta(x_u,\bm{x}_{\mathcal{N}_u})} (1-p_1)^{x_u + \bigoplus_{v\in\mathcal{N}_u}x_v}
  \right]
  \notag\\
  &\quad
  \underset{(u,v)\in E}{\widetilde{\prod}} (1-p_2)^{\theta(x_u,x_v,\bm{x}_{\tilde{\mathcal{N}}_u},\bm{x}_{\tilde{\mathcal{N}}_v})}
  ,
\end{align}
where the tilde over $\Pi$ states that the edges are introduced in a certain order.
An explicit expression for the Bell state correlators for the postselection purification protocol at first order in $p$ can be obtained from \eqref {eq:multipartite_fixed}:
\begin{equation}
  \expected{(K_{u_a})^{x_{u_a}}(K_{u_b})^{x_{u_b}}} \sim 1 - (\bar{x}_{u_a}x_{u_b}+2x_{u_a})p
  .
\end{equation}

\subsection{Bipartite B protocol}

\begin{figure}[tb]
  \begin{center}
    \includegraphics[width=.45\textwidth]{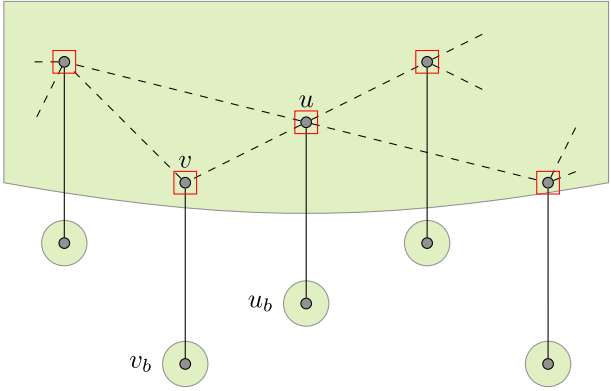}
  \end{center}
  \caption{Bipartite B protocol.
  The upper green area represents the ``central'' node, where the local graph state with the structure of the network is generated.
  The lower green circles are nodes in the network.
  Small, gray dots correspond to qubits, and lines connect neighbors.
  Dashed lines, corresponding to the edges of the local graph, are created after solid lines, corresponding to the distributed bipartite states.
  Squares in red indicate $X$ measurements used to teleport the local graph states.}
  \label{fig:channels2}
\end{figure}
The previous strategy can be improved by directly connecting the local ends of Bell pairs by means of \textsc{cphase}s, and then teleporting the local graph state performing only one $X$ measurement per node---instead of the various sources of noise induced by the Bell-measurement (\textsc{cphase} and two $X$ measurements).
The initial state is now $\bigotimes_{u\in V(G)} \rho_{g_u}$, where qubits are labeled as $u$ if they belong to the central node and $u_b$ if they are in the distributed nodes (see Fig.~\ref{fig:channels2}).
The sources of noise are: the purity of Bell pairs, \textsc{cphase}s used in the preparation of the local state and the measurement involved in each teleportation.
As in the previous case, the order of \textsc{cphase}s is important: each contributes to a noise with $(1-p_2)^{\theta(x_u,x_v,x_{u_b},x_{v_b},\bm{x}_{\tilde{\mathcal{N}}_u},\bm{x}_{\tilde{\mathcal{N}}_v})}$.
Note that the \textsc{cphase}s are performed on qubits that are already connected to nodes in the network and thus affect the distributed qubits $u_b$.
The calculation is similar as before and results in a final distributed state with correlators
\begin{align}
  \expected{\bm{K}_{\bm{x}}}
  &=   
  \prod_{u\in V} \left[
  \expected{(K_{u}^{g_u})^{x_u}(K_{u_b}^{g_u})^{\bigoplus_{v\in\mathcal{N}_u}x_v}}
  (1-p_1)^{x_u}  \right]
  \notag\\
  &
  \underset{(u,v)\in E}{\widetilde{\prod}} (1-p_2)^{\theta(x_u,x_v,\bigoplus_{w\in\mathcal{N}_u}x_w,\bigoplus_{w\in\mathcal{N}_v}x_w,\bm{x}_{\tilde{\mathcal{N}}_u},\bm{x}_{\tilde{\mathcal{N}}_v})}
  .
\end{align}

\subsection{Purify subgraph and merge}

\begin{figure}[tb]
  \begin{center}
    \includegraphics[width=.45\textwidth]{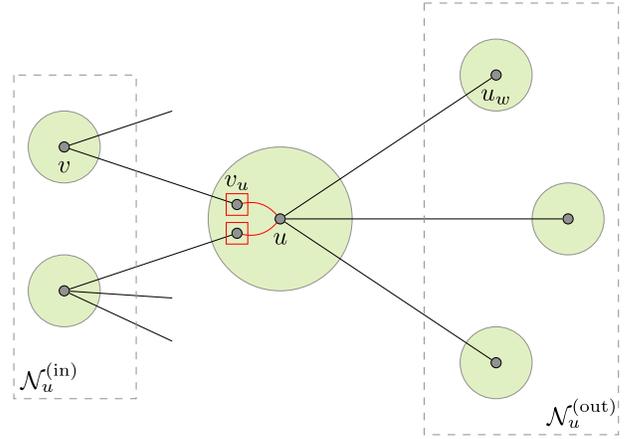}
  \end{center}
  \caption{Subgraphs purification and connection protocol.
  Green circles are nodes in the network.
  Nodes on the right are the outgoing neighborhood of $u$, while those on the left are the incoming neighborhood.
  Small, gray dots correspond to qubits, and lines connect neighbors.
  Lines and squares in red indicate \textsc{cphase}s and measurements involved in the connection of subgraphs.}
  \label{fig:subgraphs}
\end{figure}

In this strategy,  subgraph states of small size ($N$ independent) are distributed and purified, and then interconnected locally at each node to form the desired structure.

The protocol follows two steps (see Fig.~\ref{fig:subgraphs}).
To each node of degree $k$ we assign an outgoing neighborhood, with $j$ nodes, and an incoming neighborhood, with $i$ nodes ($k=j+i$).
First, each node prepares a GHZ of size $j+1$.
A GHZ graph state with $j+1$ qubits is associated to a star graph $g_u$, i.e., a graph with a central node and $j$ external nodes, called leaves.
Then, each qubit corresponding to a leaf of the GHZ is sent to one of the outgoing neighboring nodes through the depolarizing channels.
Several copies of this distributed state are created and then purified using the bicolorable graph state protocol of \cite{Dur2003}.
The final purified subgraph state with central qubit $u$ is
\begin{multline}
  \rho_u
  =\frac{1}{2^{j}}
  \sum_{x_u,\bm{x}_{\mathcal{N}_u^{\rm (out)}}}\!
  \expected{\left(K_u^{g_u}\right)^{x_u}\!\!\!\prod_{u_w\in\mathcal{N}_u^{\rm (out)}}\left(K_{u_w}^{g_u}\right)^{x_{u_w}}}
  \\
  \left(K_u^{g_u}\right)^{x_u}\!\!\!\prod_{u_w\in\mathcal{N}_u^{\rm (out)}}\left(K_{u_w}^{g_u}\right)^{x_{u_w}},
\end{multline}
where the correlators of the fixed point of the purification scheme are given in Eq.~\eqref{eq:multipartite_fixed}.
Here $u_w$ denotes the leaf qubit that has been sent to $w\in\mathcal{N}_u^{\rm (out)}(G)$.
At the same time, each node receives $i=k-j$ leaves corresponding to the GHZ states created at its incoming neighborhood.

In the second step, each node connects the central qubits of their GHZ state with the leaves they have received.
The connection is made by performing a \textsc{cphase} between the two qubits $u$ and $v_u$ and a $Y$ measurement of the received qubit $v_u$.
The effect of the $Y$ measurement is to add an edge between the two central nodes $u$ and $v$ and destroy de measured qubit $v_u$.

The action on the stabilizers can be seen in a single connection example between a central qubit $u$ of a subgraph and a leaf qubit $v_u$ of an incoming subgraph (see Fig.~\ref{fig:subgraphs}).
We focus on the stabilizer elements $(K_v^{g_v})^{x_v}(K_{v_u}^{g_v})^{x_{v_u}}(K_u^{g_u})^{x_u}$, as all the others remain unchanged.
The action of a \textsc{cphase} between $u$ and $v_u$ is $K_{v_u}^{g_v}\to K_{v_u}^{g_v}Z_{u}$ and $K_u^{g_u}\to K_u^{g_u}Z_{v_u}$.
The measurement of $Y_{v_u}$ anticommutes with the new stabilizer operators at $v$, $v_u$, and $u$, so each term changes to
\begin{multline}
  \left[ 1+(-1)^m Y_{v_u} \right]
  (K_v^{g_v})^{x_v}(K_{v_u}^{g_v}Z_u)^{x_{v_u}}(K_u^{g_u}Z_{v_u})^{x_u}
  \\
  \delta_{0,x_v\oplus x_{v_u}\oplus x_u}
  .
\end{multline}
Tracing out $v_u$ and correcting the state depending on the measurement outcome $m$ with $\exp[(-1)^m i\frac{\pi}{4}Z_{v}] \exp[(-1)^m i\frac{\pi}{4}Z_{u}]$, we obtain
\[
  (K_v\pr)^{x_v}(K_u^{g_u}Z_v)^{x_u}
  \delta_{0,x_v\oplus x_{v_u}\oplus x_u}
  ,
\]
where $K_v\pr$ is the stabilizer of $v$ with qubit $v_u$ changed to $u$.

Noise added by the measurement enters as \mbox{$(1-p_1)^{x_u\oplus x_v}$}, where we used Eq.~(\ref{eq:noise_pauli_coeff}) together with $x_{v_u}=x_u\oplus x_v$.
Noise by a \textsc{cphase} between one of the incoming leaves $v_u$ and $u$ adds $(1-p_2)^{\theta(x_u,x_v,\bm{x}_{\mathcal{N}^{\rm (out)}_u},\bm{x}_{\tilde{\mathcal{N}}^{\rm (in)}_u})}$, where $\tilde{\mathcal{N}}^{\rm (in)}_u$ is the incoming neighborhood that has already been connected to $u$.

Finally, note that if a node has $j=0$ it does not need to prepare any GHZ state.
In this case, one of the incoming leaves is used as the qubit to which all the other leaves are connected.
If there are no incoming leaves, then the node is isolated (and is thus prepared in the state stabilized by $X$).
This means that the contribution of the GHZ distribution and purification at a node $u$ is
\[
  \expected{\bm{K}_{\bm{x}}^{g_u}}
    \setlength{\arraycolsep}{1pt}
  = \left\{ \begin{array}{ll}\displaystyle
    \expected{(K_u^{g_u})^{x_u}\!\! \!\!\prod_{w\in\mathcal{N}^{\rm (out)}_u}\!\!\!\!(K_{u_w}^{g_u})^{x_u\oplus x_w}}
  & \text{if $j>0$,}
    \\
    1
  & \text{if $j=0$, $i>0$,}
    \\
    (1-p_1)^{x_u}
 & \text{if $j=i=0$.}
  \end{array} \right.
\]

All in all, the final correlators for the subgraph protocol can be written as, 
\begin{align}
\expected{\bm{K}_{\bm{x}}}
 Ê&=
 Ê\prod_{u\in V}
 Ê\left\{
 Ê\expected{\bm{K}_{\bm{x}}^{g_u}}
 Ê\prod_{v_a\in\mathcal{N}\pr_u} \left[
 Ê(1-p_1)^{x_u\oplus x_{v_a}}
 Ê\right.\right.
 Ê\notag\\
 Ê&\quad
 Ê\left.\left.
 Ê(1-p_2)^{\theta(x_u,\bm{x}_{\mathcal{N}^{\rm out}_u},x_{v_1},\dots,x_{v_a})}
 Ê\right]
 Ê\right\}.
 Ê\label{eq:expectedK}
\end{align}
Here, $\mathcal{N}\pr_u$ denotes the incoming neighborhood $\mathcal{N}_u^{\rm in}$ if $j>0$, or the incoming neighborhood minus the first incoming neighbor, $\mathcal{N}_u^{\rm in}\backslash v_1$, in the case $j=0$ where the first neighbor $v_{1}$ is used as the qubit to which all remaining edges are connected.

\subsection{\label{sec:fidelity}Fidelity}

The fidelity for a distributed graph $G$ is $F^G_N=\frac{1}{2^N}\sum_{\bm{x}}\expected{\bm{K^G}_{\bm{x}}}$, and is in general hard to calculate.
Setting $p_1=p_2=p$, all errors from the \textsc{cphase}s and measurements enter as factors
\begin{equation}
(1-p)^{h(\bm{x})} = \mathrm{e}^{ -\beta h(\bm{x})}
\label{eq:exp1mp}
\end{equation}
where we have defined $\beta=-\log(1-p)$.
To first order in $p$, the correlators of the purified graph states \eqref{eq:multipartite_fixed} can also be written in the same form,
\begin{equation}
  1-h(\bm{x})p \sim \e^{-ph(\bm{x})} \sim \e^{-\beta h(\bm{x})}
  .\label{eq:exp1mp2}
\end{equation}
Hence, the correlators of the generated graph state can be written as \mbox{$\expected{\bm{K}_{\bm{x}}}=\exp[-\beta H^G(\bm{x})]$}, where $H^G(\bm{x})$ is the sum of the different noise terms $h(\bm{x})$.
The fidelity thus resembles the partition function of a system with Hamiltonian $H^G(\bm{x})$ and inverse temperature $\beta$:
\begin{equation}
  F^G_N
  =
  \frac{1}{2^N} \sum_{\bm{x}} \e^{-\beta H^G(\bm{x})}
  .
\end{equation}

The ``Hamiltonian'' $H^G(\bm{x})$ can be expressed as the sum of many-body $n$-local Hamiltonians of the form 
\begin{align}
  \theta(x_1,x_2,\dots,x_n)
  &=
  1-\overline{x_1}\,\overline{x_2}\dots\overline{x_n}
  \label{eq:cphase_manybody}
  \\
  x_u\oplus x_v
  &=
  x_u\overline{x_v}+\overline{x_u}x_v
  \label{eq:measurement_manybody}
  \\
  \overline{x_u} \left\lceil \frac{\sum_{b=1}^j x_b}{2} \right\rceil
  &=
 \frac{1}{2}\left( \overline{x_u} {\sum_{b=1}^j x_b} + \overline{x_u}\bigoplus_{b=1}^j x_b\right)
  \label{eq:purif_manybody}
  .
\end{align}
The last term $\bigoplus_{b=1}^jx_b$ in Eq.~(\ref{eq:purif_manybody}) is a $j$-body interaction term.
 By recursively using Eq.~(\ref{eq:measurement_manybody}), this term can be seen to be equal to the sum over all index permutations of $\sum_{a \textrm{ odd} }x_1\dots x_a\overline{x_{a+1}}\dots
\overline{x_j}$.
That is, we have rephrased our problem of computing the fidelity of a distributed large graph state as that of computing the thermal properties of a classical many-body Ising-type system, where the indices $x$ take the role of classic spins.
The corresponding Hamiltonian will inherit the topology of the underlying graph and its precise expression will depend on the graph-growth protocol used.

\medskip

We are interested in the rate at which $F^G_N$ decays,
\begin{equation}
  \beta f^G_N=-\frac{1}{N}\log F^G_N,
  \label{eq:decay}
\end{equation}
where $N f^G_N$ is the analog of the free energy of the system.
A good reason to study this quantity is that in statistical systems such as a complex network (which is modeled as an ensemble of graphs, each associated to a probability---see Section \ref{sec:CN}), the partition function (fidelity) in itself is not an extensive quantity, while the free energy
is typically extensive and self-averaging (see, for example, Ref.~\cite{Amit1989} p.~188).

We can further exploit the statistical physics analogy and apply the known methods and understanding to compute the rate at which the fidelity decays for the different proposed protocols.
We are interested in a regime where the noise in the operations is low, which corresponds to the high temperature limit.
In addition, in the cases under study, each spin (or node) is effectively coupled to several spins, either as nearest- or second-nearest neighbors.
These are conditions for which mean-field approximation is very well suited:
Eqs.~(\ref{eq:cphase_manybody}) to (\ref{eq:purif_manybody}) can be linearized using the standard mean-field approximation to express the Hamiltonian as the sum of a constant term plus  linear terms in $x_u$.
For this purpose we take $x_u\to s+\delta_u$ (and $\overline{x_u}\to1-s-\delta_u$) where $s$ is the value of the mean--field and  $\delta_{u}$ are the arguably small fluctuations of $x_{u}$ around its mean value.
Keeping only the linear terms in the fluctuations, the different terms present in the Hamiltonian become,
\begin{align}
  \theta(x_1,x_2,\dots,x_n)
  &=
  1-(1-s)^n+(1-s)^{n-1}\sum_{a=1}^n\delta_a
  \\
  x_u\oplus x_v
  &=
  2s(1-s)+(1-2s)(\delta_u+\delta_v)
  \\
  \overline{x_u} \left\lceil \frac{\sum_b x_b}{2} \right\rceil
  &=
  \frac{1-s}{2}\left( js + \sum_{b=1}^j \delta_b \right) - \frac{\delta_u}{2} js
  \notag\\
  &\quad
  + \frac{1-s}{2}\left\{ \frac{1}{2} \left[1-(1-2s)^j\right] \right. 
  \notag\\
  &\quad\left.
  + (1-2s)^{j-1} \sum_{b=1}^j \delta_b \right\}
  \notag\\
  &\quad
  - \frac{\delta_u}{2} \frac{1}{2} \left[1-(1-2s)^j\right]
  .
\end{align}

With this linearization, the new mean-field Hamiltonian takes the form $H^{\rm MF} = \sum_u A_u + \sum_u B_u x_u$, where $A_u$ and $B_u$ are functions of $s$.
Hence, the sum over ${\bm{x}}$, i.e., ``configurations'', in the fidelity can now be carried out with ease,
\begin{align}
  F
  &=
  \frac{1}{2^N} \sum_{\bm{x}} \prod_{u\in V} e^{-\beta A_u} e^{-\beta Bx_u}
  \notag\\
  &=
  \prod_{u\in V} e^{-\beta A_u} e^{-\beta B/2} \left( \cosh \frac{\beta B_u}{2} \right)
  ,
\end{align}
and the decay rate becomes  $\beta f^{\rm MF}$, with
\begin{equation}
\label{eq:MFf}
  f^{\rm MF}
  =
  \frac{1}{N} \sum_u A_u + \frac{1}{N} \sum_u \frac{B_u}{2} - \frac{1}{\beta N} \sum_u \log\cosh \frac{\beta B_u}{2}
  .
\end{equation}
The value of $s$ is found by adding an artificial linear term to the Hamiltonian (playing the role of an external magnetic field), changing $B_u\to B_u+\xi$, and 
requiring consistency in the definition 
\[
  s=\expected{x_{u}}=\left. \frac{\partial f^{\rm MF}}{\partial \xi}\right|_{\xi\to 0}
  .
\]
One is hence left with the trascendental equation for $s$,
\begin{equation}
  s
  =
  \frac{1}{2} - \frac{1}{2N} \sum_u \tanh \frac{\beta B_u}{2}
  .\label{eq:slinear}
\end{equation}
Its solution $s^{*}$ can be substituted back in the expression for $ f^{\rm MF}$  \eqref{eq:MFf} to obtain the desired result.
In the cases of interest here, $B_{u}$ is some polynomial which remains bounded for all values of \mbox{$s\in[0,1]$}.
Hence, to leading order in $p$ we can approximate $\tanh \frac{\beta B_u}{2}\approx \frac{p B_u}{2}$, arriving to 
\begin{equation}
s\approx\frac{1}{2}-\frac{1}{2} \left. p B_u\right|_{s=1/2}
\end{equation}
for small enough $p$.

\section{\label{sec:nets}Network examples}

\subsection{Closed linear cluster}

We first study the case of a one-dimensional network, in which a linear cluster state is created.
This case is remarkable because we can compute the exact fidelity for any cluster size.
For symmetry, we consider a closed linear cluster state, where all nodes have degree 2.
The order of the \textsc{cphase}s in the creation of the local graph in protocols Bipartite A and B, and the size of subgraphs in protocol Subgraphs, gives different results for the fidelity.
For simplicity, in Bipartite A and B, we consider \textsc{cphase}s applied to successive nodes.
Except for the noise of the first gate (which affects only two nodes) and of the last one (which affects four nodes), all the gates contribute to the noise of three nodes.
Each of the correlators is thus
\begin{align}
  \expected{\bm{K}_{\bm{x}}}
  &\simeq
  \prod_{u\in V}
  (1-p_1)^{x_u} (1-p_1)^{x_u + x_{u-1}\oplus x_{u+1}}
  \notag\\&\qquad
  (1-p_2)^{2\theta(x_{u-1},x_u,x_{u+1})}
  \notag\\&\qquad
  \expected{(K_{u_a}^{g_u})^{x_{u-1}\oplus x_{u+1}}(K_{u_b}^{g_u})^{x_u}}
  \label{eq:Kx_linear_channels1}
\end{align}
for the Bipartite A protocol and
\begin{align}
  \expected{\bm{K}_{\bm{x}}}
  &\simeq
  \prod_{u\in V}
  (1-p_1)^{x_u}
  (1-p_2)^{\theta(x_{u-1},x_u,x_{u+1},x_{u+2})}
  \notag\\&\qquad
  \expected{(K_{u}^{g_u})^{x_u}(K_{u_b}^{g_u})^{x_{u-1}\oplus x_{u+1}}}_u
  \label{eq:Kx_linear_channels2}
\end{align}
for the Bipartite B protocol.
Recall that the correlator of the purified subgraphs is given in \eqref{eq:multipartite_fixed}.

\begin{figure}[tb]
  \begin{center}
    \includegraphics[width=.45\textwidth]{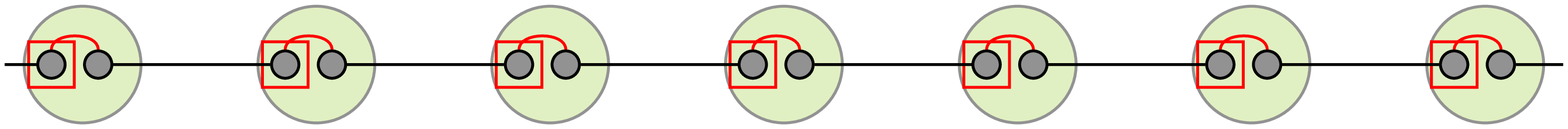}
    \includegraphics[width=.45\textwidth]{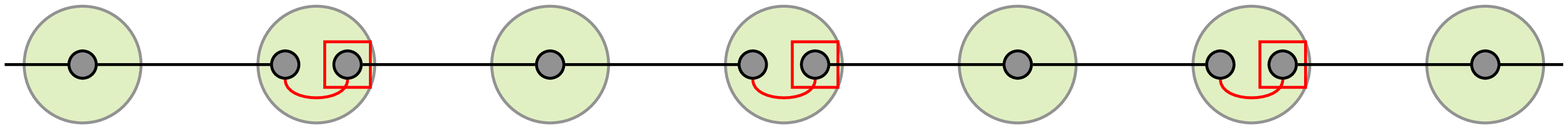}
  \end{center}
  \caption{Subgraph protocol in a linear cluster.
  Top: S1, using subgraphs of degree 1.
  Bottom: S2, using subgraphs of degree 2.}
  \label{fig:linear_subgraphs_example}
\end{figure}
The subgraph protocol, on the other hand, depends on the size of the subgraphs.
There are two extreme strategies, depicted in Fig.~\ref{fig:linear_subgraphs_example}: create subgraphs of degree 1, and connect them successively (S1), or create subgraphs of degree 2 every second node, and connect them at their common neighbor (S2).
One could also adopt an intermediate strategy, in which the degree of each GHZ is selected at random, resulting in a mix of subgraphs of degree 1 and 2 ---its performance, which will not be reported here, falls in between the two extreme strategies.
In the first case (S1), Eq.~(\ref{eq:expectedK}) reads
\begin{multline}
  \left\langle \bm{K}_{\bm{x}} \right\rangle
  =
  \prod_{u\in V}
  (1-p_2)^{\theta(x_u,x_{u+1},x_{u+2})} (1-p_1)^{x_u\oplus x_{u+1}}
  \\
  \expected{(K_u^{g_u})^{x_u} (K_{u_{u+1}}^{g_u})^{x_u\oplus x_{u+1}}},
  \label{eq:expectedK_ring_2-2}
\end{multline}
while in the second (S2) it is
\begin{multline}
  \expected{\bm{K}_{\bm{x}}}
  =
  \prod_{u\in V_{\rm even}}
  (1-p_2)^{\theta(x_u,x_{u+1},x_{u+2})} (1-p_1)^{x_{u+1}\oplus x_{u+2}}
  \\
  \expected{(K_u^{g_u})^{x_u}(K_{u_{u-1}}^{g_u})^{x_u\oplus x_{u-1}}(K_{u_{u+1}}^{g_u})^{x_{u+1}}}
  .
  \label{eq:expectedK_ring_3-0}
\end{multline}
Note that in the latter the product is over \mbox{$u=2,4,\dots\in V_{\rm even}$}.

Let us now define a domain of $\bm{x}=(x_1x_2\dots x_N)$ as a sequence $(x_ux_{u+1}\dots x_{u+l})$ where all $x_v$, for $v=u,\dots,u+l$, have the same value (either 0 or 1), and where $x_{u-1}$ and $x_{u+l+1}$ have a different value.
We can differentiate between a sequence of zeros or ones.
Taking Eq.~(\ref{eq:expectedK_ring_2-2}) as an example, one can observe that except for the term $(1-p_2)$, $\expected{\bm{K}_{\bm{x}}}$ factorizes into expected values in each component of ones, as the exponents of the noise terms can be different from zero if at least one of the bits $x$ is equal to one, and the terms are uncorrelated with those on different domains.
Considering all noise terms, Eq.~(\ref{eq:expectedK_ring_2-2}) can be expressed as a function of the total number of ones in the sequence $\bm{x}$, $n=|\bm{x}|$; the number $c_1$ of domains of only one zero; and the number of domains $c_{2^*}$ with two or more zeros:
\begin{align}
  \expected{\bm{K}_{\bm{x}}}
  &=
  (1-p_2)^{c_1+2c_{2^*}+n} (1-p_1)^{2(c_1+c_{2^*})}
  \notag\\
  &\quad
  \expected{K_{u_v}^{g_u}}^{c_1+c_{2^*}} \expected{K_u^{g_u}}^{n-c_1-c_{2^*}} \expected{K_u^{g_u}K_{u_v}^{g_u}}^{c_1+c_{2^*}}
  .
  \label{eq:expected_S1_domains}
\end{align}
The fidelity is then
\[
  F_N
  =
  \frac{1}{2^N} \sum_{c_1,c_{2^*},n} g(c_1,c_{2^*},n,N) \expected{\bm{K}_{\bm{x}}(c_1,c_{2^*},n)}
  ,
\]
where $g(c_1,c_{2^*},n,N)$ is the number of $\expected{\bm{K}_{\bm{x}}(c_1,c_{2^*},n)}$ elements with these parameters in a graph of $N$ vertices.
This sum can be computed exactly by turning to generating functions.
The function generating $g(c_1,c_{2^*},n,N)$ is $G(x,y_1,y_2,z)=\sum g(c_1,c_{2^*},n,N) x^ny_1^{c_1}y_2^{c_{2^*}}z^N$.
We derive its closed form in Appendix~\ref{sec:gf_domains}, which reads
\begin{equation}
  G
  =
  \frac{1-xz^2(1-y_1+2(y_1-y_2)z)}{1-(1+x)z+x(1-y_1)z^2+x(y_1-y_2)z^3}
  .
  \label{eq:gf_linearcluster}
\end{equation}
The fidelity for a state of size $N$ is then
\begin{equation}
  F
  =
  \frac{1}{2^N}\frac{1}{N!}\partial_z^N G(x,y_1,y_2,z)
  ,
  \label{eq:fidelity_from_gf}
\end{equation}
evaluated at
\begin{align}
  x   & = (1-p_2)                                      \expected{K_u^{g_u}}                                        ,\label{eq:gfS1_x}\\
  y_1 & = (1-p_2)   (1-p_1)^2 \expected{K_{u_v}^{g_u}} \expected{K_u^{g_u}}^{-1} \expected{K_u^{g_u}K_{u_v}^{g_u}} ,\label{eq:gfS1_y1}\\
  y_2 & = (1-p_2)^2 (1-p_1)^2 \expected{K_{u_v}^{g_u}} \expected{K_u^{g_u}}^{-1}\expected{K_u^{g_u}K_{u_v}^{g_u}}  ,\label{eq:gfS1_y2}
\end{align}
and $z=0$.
The factorization of $\expected{\bm{K_{\bm{x}}}}$ for Subgraphs S2 and Bipartite protocols depend on slightly different parameters, and the corresponding generating functions can also be computed (see Appendix~\ref{sec:gf_domains}).

\begin{figure}[tb]
  \begin{center}
    \includegraphics[width=.45\textwidth]{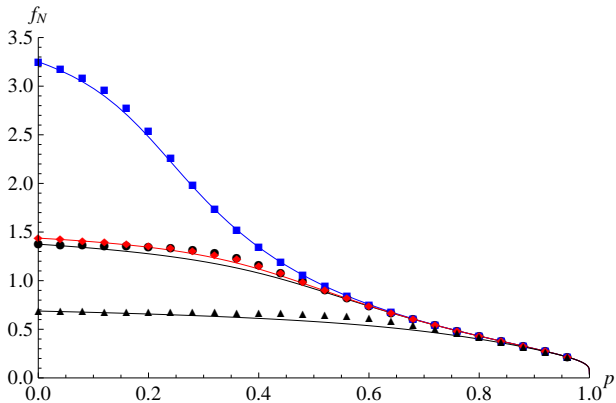}
  \end{center}
  \caption{Rescaled decay rate $f_N$ of a closed linear cluster of $N=10$ nodes with perfect purification.
  Blue (squares): Bipartite A; red (diamonds): Bipartite B; black: Subgraphs S1 (circles) and S2 (triangles).
  Dots correspond to the mean-field approximation, lines to the exact result using generating functions.}
  \label{fig:f_linear_n10_MF}
\end{figure}

We will first consider the ideal case of perfect purification, and later tackle the noisy case. In the ideal scenario,  we can substitute all subgraph correlators by $1$ and Eq.~(\ref{eq:fidelity_from_gf}) gives the exact fidelity for all $p$. The corresponding decay exponent $f_N$ is plotted in Fig.~\ref{fig:f_linear_n10_MF}.

We now compare these exact results with the mean-field approximation.
As we introduced in the previous section, $\expected{\bm{K}_{\bm{x}}}\sim\exp[-\beta H(\bm{x})]$.
For the Subgraph S1 protocol, $H(\bm{x})=\sum_{u\in V}h_u(\bm{x})$, with
\begin{equation}
  h_u(\bm{x})
  =
  \theta(x_u,x_{u+1},x_{u+2}) + x_u\oplus x_{u+1}
  .
  \label{eq:hu_linear_S1}
\end{equation}
Recall that we are still considering perfect purification.
Substituting $x_u\to s+\delta_u$ and keeping only linear terms in $\delta_u$, this takes the form of $h_u^{\rm MF}(\bm{x})=a_u + b_u x_u + c_u x_{u+1} + d_u x_{u+2}$, with
\begin{align}
  a_u
  &=
  (5-2s) s^2
  \\
  b_u
  &=
  2 - 4s + s^2
  \\
  c_u
  &=
  2 - 4s + s^2
  \\
  d_u
  &=
  (1-s)^2
  .
\end{align}
All qubits $u$ are equivalent, so the coefficients $a,b,c,d$ are the same for every node and the total Hamiltonian is $H^{\rm MF}=AN+B\sum_ux_u$, with $A=a$ and $B=b+c+d$.
Hence, the fidelity in the mean-field approximation is
\begin{equation}
  F_N^{\rm MF}
  =
  \e^{-\beta AN} \e^{-\beta BN/2} \left( \cosh \frac{\beta B}{2} \right)^N
  ,
\end{equation}
and the decay rate
\begin{equation}
  f^{\rm MF}
  =
  A + \frac{B}{2} - \frac{1}{\beta} \log \cosh \frac{\beta B}{2}
  .
\end{equation}

The same approximation can be made for the other protocols.
The local Hamiltonian of Subgraphs S2 is
\begin{equation}
  h_u(\bm{x})
  =
  \theta(x_u,x_{u+1},x_{u+2}) + x_{u+1}\oplus x_{u+2}
  .
  \label{eq:hu_linear_S2}
\end{equation}
Note that, in this case, the total Hamiltonian is the sum for $u$ even.
In the case of Bipartite A and B, the local Hamiltonians are
\begin{align}
  h_u(\bm{x})
  &=
  2x_u + x_{u-1}\oplus x_{u+1} + 2\theta(x_{u-1},x_u,x_{u+1})
  \label{eq:hu_linear_channels1}
\intertext{and}
  h_u(\bm{x})
  &=
  x_u+\theta(x_{u-1},x_u,x_{u+1},x_{u+2})
  ,
  \label{eq:hu_linear_channels2}
\end{align}
respectively.
The mean-field results are shown in Fig.~\ref{fig:f_linear_n10_MF} together with the exact result.
The agreement is remarkably good, specially considering that this is a one-dimensional network configuration. We observe that the decay rate has a linear dependence in $p$ but soon higher order (non-linear) terms start to kick in. Here, we have studied and plotted the solution for a wide range of $p$. This has been done for completeness and to check the validity of the mean-field approximation, but recall that the aim in this paper is solely to compute the decay rate to leading order in $p$. The reason being that  in realistic scenarios $p$ will be strongly limited by threshold values required for subgraph purification.

We now consider the realistic scenario following the same procedure as before but taking into account the corrections due the noisy purified subgraphs. This can be done by approximating the correlators  by their linear correction around unity \eqref{eq:multipartite_fixed}. Consequently, an additional term \eqref{eq:exp1mp2} must be added to the local Hamiltonians (\ref{eq:hu_linear_S1}), (\ref{eq:hu_linear_S2}), (\ref{eq:hu_linear_channels1}), (\ref{eq:hu_linear_channels2}) for each of the different protocols. It is important to bear in mind that all subsequent results have to be taken consistently up to leading order in $p$. This leading order is the one that enters in the features that we seek: the linear dependence  of the fidelity decay rate around $p\sim 0$ (for arbitrarily large $N$).
From equation \eqref{eq:MFf} and \eqref{eq:slinear}  it is immediate to see that at this order the whole contribution to the decay rate is given by the first two terms in \eqref{eq:MFf} evaluated at $s=1/2$, i.e.,
\begin{equation}
  f_N  =
  \left.
  A + \frac{B}{2}
  \right|_{s=1/2}
  =
  \frac{1}{N2^N} \sum_{\bm{x}} H(\bm{x})
  .
  \label{eq:linear_fp_1storder}
\end{equation}
The second equality is straightforward to get from the definitions of $A$ and $B$, and it states that in this regime (low $p$) the decay rate is dominated by the exponents of the typical  sequences  ${\bm{x}}$. The expected value of $H(\bm{x})$  (over sequences $\bm{x}$) can be easily carried out since $x_u$ are independent variables of mean $1/2$ and it leads to fidelities that decay as $p f_N N$ with  $f_N$ equal to $9/2=4.5$, $43/16\approx2.7$, $21/8\approx2.6$, and $13/8\approx1.6$ for the Bipartite A and B and Subgraphs S1 and S2, respectively.

\begin{figure}[tb]
  \begin{center}
    \includegraphics[width=.45\textwidth]{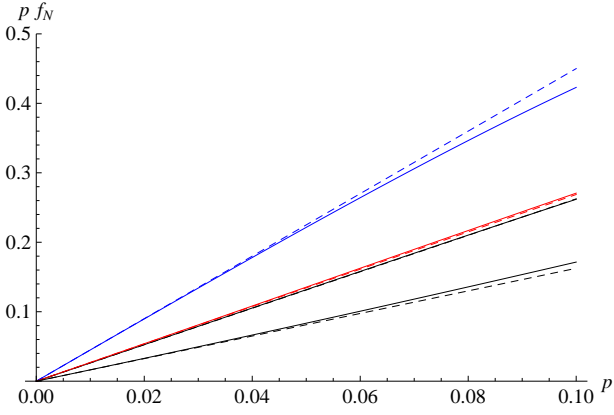}
  \end{center}
  \caption{Fidelity decay rate $pf_N$ of a closed linear cluster of $N=100$ nodes.
  From upper to lower lines: Bipartite A (blue), Bipartite B (red), Subgraph S1 (black) and Subgraph S2 (black).
  Solid lines correspond to the generating function result, using the first order approximation of the purification scheme of Eq.~(\ref{eq:multipartite_fixed}).
  Dashed lines are the first-order result of Eq.~(\ref{eq:linear_fp_1storder}).
  Dots correspond to the mean-field approximation, dashed lines to the fidelity at $p\to0$.}
  \label{fig:fp_linear_n100_1storder}
\end{figure}

These coincide with the decay rates obtained by the generating function method in the same limit, where in Eq.~(\ref{eq:gfS1_x})--(\ref{eq:gfS1_y2}) we substitute $\expected{K_u^{g_u}} = \expected{K_u^{g_u}K_{u_v}^{g_u}} = 1-2p$ and $\expected{K_{u_v}^{g_u}}^{c_1+c_{2^*}} = 1-p$.
The results (see Fig.~\ref{fig:fp_linear_n100_1storder}) show that the subgraph protocols provide output fidelities comparable, if not better, than those given by the protocols based on channel purification.
This is remarkable keeping in mind that purifying channels is much more demanding in terms of resources and efficiency. 

As we mentioned at the beginning of the section, the order of the \textsc{cphase}s in the Bipartite protocols can give different results, but the decay rates do not change much.
For example, if the \textsc{chpase}s are first applied to every second edge, and then to the remaining edges, the decay rate is $143/32\approx 4.5$ for Bipartite A, while it does not change for Bipartite B.

\subsection{\label{sec:CN}Complex networks}

Now we study the behavior of the protocols in complex networks.
Complex networks are characterized by statistical properties, and can be modeled as an ensemble of graphs $\mathcal{G}$, with a probability $P(G)$ assigned to every graph $G$ in the ensemble.
A property $O$ of a complex network is defined as its average over the ensemble, $\overline{O}=\sum_{G\in\mathcal{G}}P(G)O^G$ (note that the overline here stands for the ensemble average).
Some of these properties are \emph{self-averaging}, meaning that for large systems (in the limit $N\to\infty$) a property of a given graph realization $G$ is the same as the average over different realizations of $G\in\mathcal{G}$ \cite{Mezard1987,Amit1989}.
As we discussed in Section~\ref{sec:fidelity}, the free energy is typically self-averaging, and is the one we use here as a figure of merit.

We consider random networks with uncorrelated degree distribution $p_k$, which defines the ensemble.
This family of networks includes the Erd\H os-R\'enyi model (when $p_k$ is a Poisson distribution) as well as exponential and scale-free networks \cite{Newman2001_random}.
It is useful to express the degree distribution by its generating function $g_p(z)=\sum_kp_kz^k$.
Related to $p_k$ is the probability that, following a random edge, we arrive at a vertex with $k$ other edges.
This probability is $r_k=(k+1)p_{k+1}/\expected{k}$, and $k$ is called the \emph{excess degree}.
It is generated by $g_r(x)=g_p\pr(x)/\expected{k}$.
The edges in the network are undirected, but the creation of the graph state via the Subgraph protocol is ``directed-like'', as each node can have incoming and outgoing neighbors.
Thus, we also consider the degree distribution $p_{i,j}$, where $i$ is the in-degree and $j$ the out-degree.
This distribution defines the implementation of the protocol.
For example, in the linear cluster, S1 was defined by $p_{1,1}=1$ for all nodes, and S2 by $p_{0,2}=1$ and $p_{2,0}=1$ at even and odd nodes, respectively.
The distribution $p_{i,j}$ is constrained by $\sum_{i,j} (j-i)p_{i,j}=0$, which means that $\expected{i}=\expected{j}=\expected{k}/2$, but note that, in general, it can be correlated ($p_{i,j}\neq p_ip_j$).
The function that generates $p_{i,j}$ is $g_p(x,y)=\sum_{i,j}p_{i,j}x^iy^j$.
\medskip

In the case of complex networks, we are interested in the average $\overline{f_N}=\sum_{G\in\mathcal{G}}P(G)f^G_N$ over the ensemble.
We expect that, for large $N$, $f_N$ goes to $\overline{f}$ due to self-averaging.
In order to average $f_N$ we have to compute
\[
  \overline{f_N}
  =
  -\frac{1}{\beta N}\sum_{G\in\mathcal{G}}P(G)\log\frac{1}{2^N}\sum_{\bm{x}}e^{-\beta H^G(\bm{x})},
\]
which corresponds to a quenched average, where the disorder corresponding to the network topology, given by $P(G)$, is frozen with respect to that of the correlation operator index, $\bm{x}$.
Computing this quantity is in general an extremely challenging problem, specially in our case that we have structured disorder and k-body interaction between the spins. 
Nevertheless, we are interested in the particular regime of high temperatures (low $p$) far away from critical phenomena, long range correlations, and other difficulties that appear at low temperatures.
In this regime we can directly use  (\ref{eq:linear_fp_1storder}):
\begin{equation}
  \overline{f}
  =
  \frac{1}{N}\sum_{G\in\mathcal{G}} P(G) \frac{1}{2^N}\sum_{\bm{x}} H^G(\bm{x})
  =
  \frac{1}{N}\overline{\frac{1}{2^N}\sum_{\bm{x}} H^G(\bm{x})}
  .
\end{equation}

The Hamiltonians for the Bipartite protocols are of the form
\begin{equation}
  H^G(\bm{x})
  =
  \sum_{u\in V}h_u(\bm{x}) + \underset{(u,v)\in E}{\widetilde{\sum}} h_{u,v}(\bm{x})
  .
\end{equation}
In the Bipartite A, the local Hamiltonians are
\begin{align}
  h_u(\bm{x})
  &=
  x_u + \overline{\bigoplus_{v\in\mathcal{N}_u} x_v} \cdot x_u + 2 \bigoplus_{v\in\mathcal{N}_u} x_v\nonumber\\
  &\quad
  + \theta(x_u,\bm{x}_{\mathcal{N}_u}) + x_u + \bigoplus_{v\in\mathcal{N}_u} x_v
  \intertext{and}
  h_{u,v}(\bm{x})
  &=
  \theta(x_u,x_v,x_{\tilde{\mathcal{N}}_u},x_{\tilde{\mathcal{N}}_v})
  ,
\end{align}
while in the Bipartite B,
\begin{align}
  h_u(\bm{x})
  &=
  \overline{x_u} \bigoplus_{v\in\mathcal{N}_u}x_v + 2 x_u + x_u
  \\
  \intertext{and}
  h_{u,v}(\bm{x})
  &=
  \theta\left(x_u,x_v,\bigoplus_{w\in\mathcal{N}_u} x_w,\bigoplus_{w\in\mathcal{N}_v} x_w,x_{\tilde{\mathcal{N}}_u},x_{\tilde{\mathcal{N}}_v}\right)
  .
\end{align}
As before, the tilde over the summatory and the neighborhood stands for the order in which \textsc{cphase}s are performed in the local graph.
We can substitute it by the expected effect of an edge, which depends on the number of edges already connected to nodes $u$ and $v$.
This number is $n_u+n_v$ with probability $\frac{1}{2^{k_u+k_v}}{k_u\choose n_u}{k_v\choose n_v}$.
Here, $k_u$ and $k_v$ are the excess degrees of the vertices in edge $(u,v)$, so the average of the network ensemble has to be performed using probabilities $r_{k_u}$ and $r_{k_v}$:
\begin{equation}
  \overline{h_{u,v}(\bm{x})}
  =
  \frac{1}{2^N} \sum_{\bm{x}} \sum_{k_uk_v}r_{k_u}r_{k_v} h_{u,v}(\bm{x})
  .
\end{equation}
In the Bipartite A, the average effect of each of these edges is $1-\frac{1}{4}\left[ g_r(3/4) \right]^2$.
In the Bipartite B, it is $1-\frac{1}{16}\left[ g_r(1/2)+g_r(3/4) \right]^2$.
The terms in $h_u(\bm{x})$, on the other hand, depend directly on the degree, and the average is performed over $p_k$:
\begin{equation}
  \overline{h_u(\bm{x})}
  =
  \frac{1}{2^N} \sum_{\bm{x}} \sum_{k}p_k h_u(\bm{x})
  .
\end{equation}
Considering all these terms, and that the summation over $V$ contains $N$ elements, while that over $E$ contains $\expected{k}N/2$, the decay rates are
\begin{align}
  \overline{f_N}
  &=
  \frac{15}{4} - \frac{5}{4} g_p(0) - \frac{1}{2} g_p(1/2) + \frac{\expected{k}}{2} \left( 1 - \frac{1}{4} \left[ g_r(3/4) \right]^2 \right)
  \label{eq:f_cn_BipA}
\intertext{for Bipartite A and}
  \overline{f_N}
  &=
  \frac{7}{4} - \frac{1}{4} g_p(0) + \frac{\expected{k}}{2} \left( 1 - \frac{1}{16} \left[ g_r(3/4)+g_r(1/4) \right]^2 \right)
  \label{eq:f_cn_BipB}
  ,
\end{align}
for Bipartite B.

In the Subgraph protocol, where $H^G(\bm{x})=\sum_{u\in V}h_u(\bm{x})$, the local Hamiltonian is
\begin{align}
  h_u(\bm{x})
  &=
  \left[ \overline{x_u} \left\lceil \frac{\sum_{w\in\mathcal{N}_u^{\rm out}}x_w}{2} \right\rceil \right.
  \notag\\
  &\quad
  \left.
  + x_u (j^{(u)}+1) \right] (1-\delta_{0,j^{(u)}}) + x_u \delta_{0,i^{(u)}}\delta_{0,j^{(u)}}
  \notag\\
  &\quad
  + \sum_{v_a\in\mathcal{N}_u\pr} \left[
  \theta(x_u,\bm{x}_{\mathcal{N}_u^{\rm out}},x_{v_1},\dots,x_{v_a}) + x_u\oplus x_{v_a}
  \right]
  .
  \label{eq:f_cn_Sub}
\end{align}
Here $\mathcal{N}_u\pr$ is defined as in Eq.~(\ref{eq:expectedK}).
In this case, to average the Hamiltonian we have to take into account the directed degree probability $p_{ij}$, $\overline{h(\bm{x})}=\sum_{ij}p_{ij}h(\bm{x})$, giving
\begin{multline}
  \overline{f_N}
  =
  \frac{5}{8} + \frac{17\expected{k}}{16} + \frac{7}{4} g_p(0,0) - \frac{15}{8} g_p(1,0)
  \\
  - \frac{1}{2} \left[ g_p(1,1/2) - g_p(1/2,1/2) \right]
  .
\end{multline}

\begin{figure}[tb]
  \begin{center}
    \includegraphics[width=.45\textwidth]{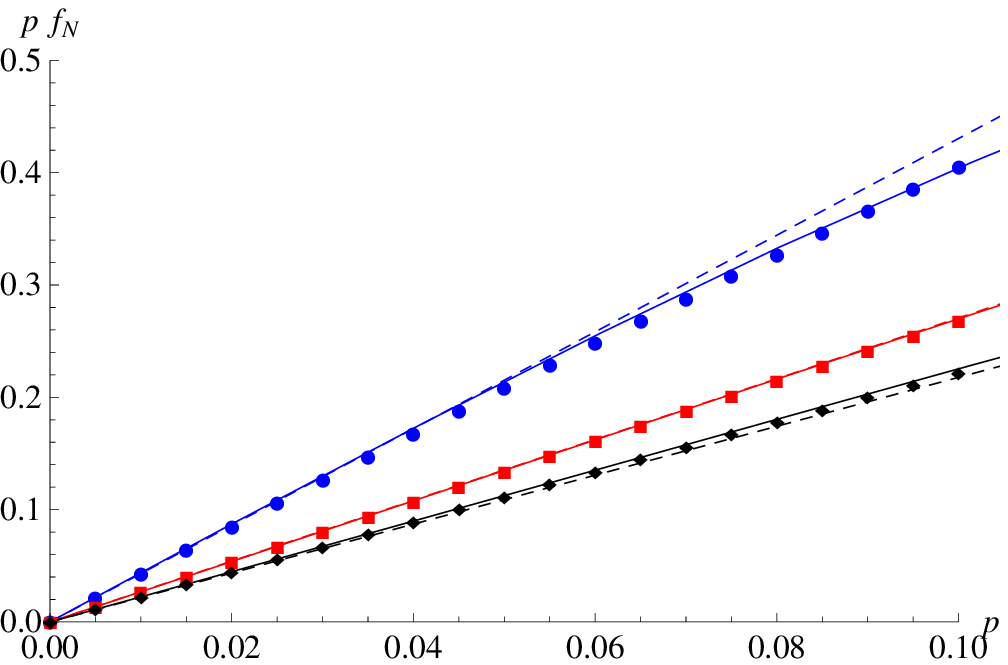}
    \includegraphics[width=.45\textwidth]{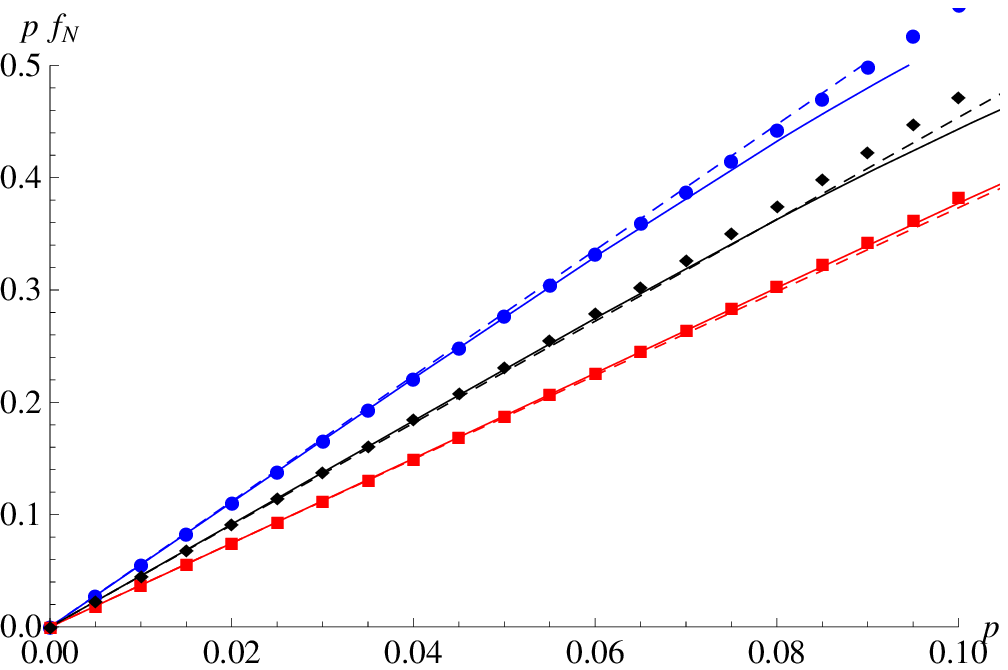}
  \end{center}
  \caption{Fidelity decay rate $p\overline{f_N}$ in an Erd\H os-R\'enyi network of mean degree $\expected{k}=2$ (top) and $4$ (bottom).
  Bipartite A (blue, circles), Bipartite B (red, squares), and Subgraphs (black, diamonds).
  Bipartite A (upper, blue), Subgraphs (middle, black) and Bipartite B (lower, red).
  Dots are the average over 10 random realizations of networks with $N=100$, randomly sampling $1\,000$ configurations of $\bm{x}$.
  Dashed lines are the results at first order in $p$, Eqs.~(\ref{eq:f_cn_BipA}), (\ref{eq:f_cn_BipB}), and (\ref{eq:f_cn_Sub}).}
  \label{fig:fp_er_p0}
\end{figure}

These results are valid for any network with uncorrelated degree distribution $p_k$.
To compare the behavior of the three protocols we consider the Erd\H os-R\'enyi model \cite{Gilbert1959,Erdos1959,Erdos1960}, which is the maximally random graph under the only constrain that the mean degree $\expected{k}$ is fixed.
The degrees in this model follow a Poisson distribution, so $g_p(x)=g_r(x)=e^{\expected{k}(x-1)}$, and for $\expected{k}>1$ there is a giant connected component of size comparable to the size of the network.
The degree distribution $p_k$ is fixed by the network model, but in the Subgraph protocol one can play with the distribution $p_{i,j}$ (as long as $\sum_{i,j}p_{i,j}\delta_{i+j,k}=p_k$).
For simplicity, here we consider that the direction of each edge is selected at random,
\begin{equation}
  p_{i,j}
  =
  \frac{p_{i+j}}{2^{i+j}}{i+j\choose i},
\end{equation}
and hence $g_p(x,y)=e^{\expected{k}\left(\frac{x+y}{2}-1\right)}$.
Fig.~\ref{fig:fp_er_p0} plots $p\overline{f_N}$ for the three protocols, showing that the best protocol depends on the mean degree of the network.
It also shows an average over 10 random realizations of an Erd\H os-R\'enyi graph of $N=100$, approximating the fidelity by the average of a random sample of $1\,000$ configurations of $\bm{x}$, $F\sim\frac{1}{1000} \sum_{\bm{x}_{\rm sample}} \expected{\bm{K}_{\bm{x}_{\rm sample}}}$.
This approximation is valid in the low $p$ regime, where the fidelity is dominated by the typical values of $\bm{x}$.
We observe that, in the Erd\H os-R\'enyi, the Subgraph protocol with random edge direction give a better (lower) decay rate $\overline{f_N}$ for $\expected{k}<2.8$, while above that the Bipartite B beats it.

\section{\label{sec:conc}Conclusions}

We have proposed a protocol to generate a graph state that spans a distributed network of any topology, in the presence of noise in the communication channels and in the local quantum operations performed in the network nodes. The protocol distributes and purifies small subgraphs, that are then merged to mimic the network structure. This allows the use of multipartite purification for small systems (that depend on the degree of the nodes, rather than on the size of the network), which makes the protocol scale efficiently. We have compared this protocol, which uses multipartite state purification and graph merging, to two other protocols that rely on channel purification (quantum repeaters). 

We have benchmarked the protocols using as a figure of merit the fidelity of the generated graph state, or, more precisely, its decay rate, as the size of the network increases.
Using generating function methods we have been able to compute exactly the fidelity for linear clusters of arbitrary size, allowing for a direct comparison with the approximate methods we develop.
We have rephrased the problem of computing the effects of noise in the operations in terms of the thermal properties of a classical spin system, with the same interaction patterns as the underlying graph.  Indeed, the fidelity itself can be seen as the analog of the partition function of such system, while its decay rate plays the role of the free energy. The well-known methods from statistical physics, such as the mean-field approximation, can be used to study its behavior.

We have also studied the three  protocols in networks with a complex structure. These complex networks are present in many real-world systems, and are of particular importance in the description of communication networks. The subgraph protocol is specially motivated to be implementable in networks of this kind, as it is highly adaptable to any network topology and it only requires the nodes to have local knowledge of the network.

Our results  shows that the protocol using subgraph purification and merging is comparable, and in some cases even better, than those which rely on bipartite states. This is quite remarkable, as the latter rely on quantum repeaters and require much more resources.  
In the complex networks case, the multipartite protocol could  be possibly  enhanced by devising an optimized strategy via the directed degree distribution $p_{i,j}$. As an example, in networks with many nodes of degree 1, one should go for a directed degree distribution in which those nodes are as much as possible the receivers of GHZ states, instead of the senders, so less connections would need to be made.
In other words, one should tend to $p_{1,0}$ as high as possible (compared to $p_{0,1}$). Also, in nodes of too high degree, the protocol might fail due to the noise threshold in the multipartite purification, which depends on the size of the GHZs (and thus, on the degree of the network nodes). In this case, one could always separate the node in two (or more) and treat them as independent nodes, each creating and distributing a subgraph among a subset of the neighborhood. This might prove useful in networks with a scale-free degree distribution, which have a long tail and a high presence of hubs.

It remains an open question to relate the decay in fidelity with the actual use one can make of graph states.
Clearly, there is a regime where the fidelity is exponentially small, but a finite decay rate still signatures valuable resources. It is important to emphasize that for the three protocols, as well as for all the networks under study, we have not only computed this decay rate, but also completely characterized the noise graph states that emerge form the protocol. For a particular application, having full knowledge of the generated graph gives probably a closer picture of the generated resource.

\begin{acknowledgments}
  We acknowledge financial support from ERDF: European Regional Development Fund.
  This research was supported by the Spanish MICINN, through contract FIS2008-01236 and through FPU grant AP2008-03048 (M.~C.); and the Generalitat de Catalunya CIRIT, contract 2009SGR-0985.
\end{acknowledgments}

\bibliography{gs_net}

\appendix

\section{\label{sec:multipartite_protocol}Multipartite purification protocol}

The recursive protocol \cite{Aschauer2005} is an entanglement purification
protocol for multipartite, two-colorable graph states, which has also been
extended to general graph states \cite{Kruszynska2006}.
It operates on two identical copies
\[
  \rho_1 \otimes\rho_2
  =
  \frac{1}{2^{2N}}
  \sum_{\bm{x}_1,\bm{x}_2} \expected{\bm{K}^{(1)}_{\bm{x}_1}} \expected {\bm{K}^{(2)}_{\bm{x}_2}}
  \bm{K}^{(1)}_{\bm{x}_1} \bm{K}^{(2)}_{\bm{x}_2}
\]
of a two-colorable graph state of size $N$ (with colors $A$ and $B$), and consists of two subprotocols (P1 and P2), each of which purifies one of the two colors.
Here $\bm{K}_{\bm{x_i}}^{(i)}$ is a stabilizer element of state $\rho_i$.
In each subprotocol, information about $\rho_1$ is transferred to $\rho_2$.
Then, $\rho_1$ is kept or discarded depending on the outcomes of measurements on $\rho_2$.
In P1, a $\textsc{cnot}_{2\to1}$ is applied to every node in $A$, and a $\textsc{cnot}_{1\to2}$ to every node in $B$.
This transforms the state to
\[
  \frac{1}{2^{2N}}
  \sum_{\bm{x}_1,\bm{x}_2}
  \expected{\bm{K}^{(1)}_{\bm{x}_1}} \expected {\bm{K}^{(2)}_{\bm{x}_2}}
  \bm{K}^{(A_1)}_{\bm{a}_1\oplus\bm{a}_2} \bm{K}^{(B_1)}_{\bm{b}_1}
  \bm{K}^{(A_2)}_{\bm{a}_2} \bm{K}^{(B_2)}_{\bm{b}_2\oplus\bm{b}_1}
  ,
\]
where $\bm{a}$ and $\bm{b}$ are the elements of $\bm{x}$ that correspond to colors $A$ and $B$. The stabilizer $K_{\bm{a}_1\oplus\bm{a}_2}^{(A_1)}$ corresponds to qubits of $\rho_1$ in color $A$, and the modulo 2 summation of the index $\bm{a}_1\oplus\bm{a}_2$ is made elementwise.
To keep notation short, we contract $\bm{K}_{\bm{a}_1\oplus\bm{a}_2}^{(A_1)}\bm{K}_{\bm{b}_2}^{(B_1)}=\bm{K}_{\bm{a}_1\oplus\bm{a}_2,\bm{b}_2}^{(1)}$, where the first subindex corresponds to color $A$ and the second to color $B$.
Then, every node in $\rho_2$ is measured: $X$ in nodes of color $A$ (outcomes $\xi$) and $Z$ in nodes of color $B$ (outcomes $\zeta$).
This gives the (unnormalized) state
\begin{multline*}
  \sum_{\bm{x}_1,\bm{x}_2}
  (-1)^{(\xi\oplus\bigoplus\zeta)\cdot\bm{a_2}}
  \expected{\bm{K}^{(1)}_{\bm{x}_1}} \expected{\bm{K}^{(2)}_{\bm{x}_2}}
  \bm{K}^{(1)}_{\bm{a}_1\oplus\bm{a}_2,\bm{b}_1}
  \delta_{\bm{0},\bm{b}_2\oplus\bm{b}_1}
  \\
  =
  \sum_{\bm{x}_1,\bm{a}_2}
  (-1)^{(\xi\oplus\bigoplus\zeta)\cdot\bm{a_2}}
  \expected{\bm{K}^{(1)}_{\bm{a}_1,\bm{b}_1}}
  \expected{\bm{K}^{(2)}_{\bm{a}_2,\bm{b}_1}}
  \bm{K}^{(1)}_{\bm{a}_1\oplus\bm{a}_2,\bm{b}_1}
  ,
\end{multline*}
where $\xi\oplus\bigoplus\zeta$ stands for $\xi_u\oplus\bigoplus_{v\in\mathcal{N}_u}\zeta_v$ for all $u\in A$.
The state is selected if $\xi\oplus\bigoplus\zeta=0$.
Summing over all the possible outcomes, the final (post-selected) state after P1 is
\[
  \frac{1}{2^{N|A|}}
  \sum_{\bm{x}_1,\bm{a}_2}
  \expected{\bm{K}^{(1)}_{\bm{a}_1} \bm{K}^{(1)}_{\bm{b}_1}}_1
  \expected{\bm{K}^{(2)}_{\bm{a}_2} \bm{K}^{(2)}_{\bm{b}_1}}_2
  \bm{K}^{(A_1)}_{\bm{a}_1\oplus\bm{a}_2} \bm{K}^{(B_1)}_{\bm{b}_1}
  .
\]
P2 is equivalent, with colors A and B interchanged.

For our protocol, we need the fixed point for a GHZ of size $j+1$, with a central node colored as $A$ and $j$ leaves colored as $B$ ($|A|=1$ and $|B|=j$).
Noise can come from \textsc{cnot}s as $1-p_2$ and from measurements in state $\rho_2$ as $1-p_1$:
\begin{align*}
  \expected{K_a \bm{K}_{\bm{b}}}^{(\rm P1)}
  &=
  \frac{1}{2} \sum_{a_2=0}^1
  \expected{K_{a\oplus a_2}^{(1)} \bm{K}_{\bm{b}}^{(1)}}
  \expected{K_{a_2}^{(2)} \bm{K}_{\bm{b}}^{(2)}}
  \\ &
  (1-p_2)^{\theta(a,a_2,\bigoplus_{b\in B} b)} \prod_{b\in B} (1-p_2)^{\theta(a,a_2,b)}
  \\ &
  (1-p_1)^{a_2} \prod_{b\in B} (1-p_1)^{a_2}
  \\
  \expected{K_a \bm{K}_{\bm{b}}}^{(\rm P2)}
  &=
  \frac{1}{2^d} \sum_{\bm{b_2}=\bm{0}}^{\bm{1}}
  \expected{K_a^{(1)} \bm{K}_{\bm{b}\oplus\bm{b_2}}^{(1)}}
  \expected{K_a^{(2)} \bm{K}_{\bm{b_2}}^{(2)}}
  \\ &
  (1-p_2)^{\theta(a,\bigoplus_{b\in B} b_2,\bigoplus_{b\in B} b)} \prod_{b\in B} (1-p_2)^{\theta(a,\bm{b}_2,b)}
  \\ &
  (1-p_1)^{\bigoplus_{b\in B} b_2} \prod_{b\in B} (1-p_1)^{b_2}
  .
\end{align*}
We now consider $p_1=p_2=p$ and approximate $\expected{K_a\bm{K}_{\bm{b}}}$ at first order in $p$.
Let the unnormalized $\expected{K_a\bm{K}_{\bm{b}}}^{\rm(P1)}\sim1-\tilde{\beta}_{a,|\bm{b}|}p$ and $\expected{K_a\bm{K}_{\bm{b}}}^{\rm(P2)}\sim1-\tilde{\alpha}_{a,|\bm{b}|}p$.
Composing P1 and P2 we can find the fixed point at first order in $p$.
In P1, each $\tilde{\beta}_{a,|\bm{b}|}$ equals $\tilde{\alpha}_{0,|\bm{b}|}+\tilde{\alpha}_{1,|\bm{b}|}$ plus a constant term:
\begin{align*}
  \tilde{\beta}_{0,|\bm{b}|}
  &=
  \alpha_{0,|\bm{b}|} + \alpha_{1,|\bm{b}|} + j+1 + \left\lceil \frac{|\bm{b}|}{2} \right\rceil,
  \\
  \tilde{\beta}_{1,|\bm{b}|}
  &=
  \alpha_{0,|\bm{b}|} + \alpha_{1,|\bm{b}|} + \frac{3}{2}(j+1)
  .
\end{align*}
These terms are normalized dividing them by $\expected{\openone}=\expected{K_0\bm{K}_{\bm{0}}}$, so the normalized first-order coefficients for P1 read $\beta_{a,|\bm{b}|}=\tilde{\beta}_{a,|\bm{b}|}-\tilde{\beta}_{0,0}$.
Similarly, for P2:
\begin{align*}
  \tilde{\alpha}_{0,|\bm{b}|}
  &=
  \frac{1}{2^j}\sum_{\bm{b}_2} (\beta_{0,\bm{b}\oplus\bm{b}_2} + \beta_{0,\bm{b}_2}) + j+1 + \left\lceil\frac{|\bm{b}|}{2}\right\rceil
  \\
  \tilde{\alpha}_{1,|\bm{b}|}
  &=
  \frac{1}{2^j}\sum_{\bm{b}_2} (\beta_{1,\bm{b}\oplus\bm{b}_2} + \beta_{1,\bm{b}_2}) + \frac{3}{2}(j+1)
\end{align*}
After normalization (dividing by $\expected{K_0\bm{K}_{\bm{0}}}$),
\begin{align*}
  \expected{K_0 \bm{K}_{\bm{b}}}
  &\sim
  1 - \left\lceil \frac{|\bm{b}|}{2} \right\rceil p,
  \\
  \expected{K_1 \bm{K}_{\bm{b}}}
  &\sim
  1 - (j+1) p
  .
\end{align*}
The fidelity is
\begin{align*}
  F
  &\sim
  1 - \frac{1}{2^{j+1}} \sum_{b=0}^j {j \choose b} \left[ \left\lceil \frac{b}{2} \right\rceil + (j+1) \right] p
  \\
  &=
  1 - \frac{5}{8} (j+1) p
  .
\end{align*}

\section{\label{sec:gf_domains}Generating functions for the domains in the closed linear cluster state}

In this appendix we derive the generating functions from which the fidelity of a closed linear cluster can be derived using Eq.~(\ref{eq:fidelity_from_gf}).
In all cases, errors from \textsc{cphase}s, measurements and multipartite purification are tracked by error parameters $p_2$, $p_1$ and $p$, respectively.

\subsection{Subgraphs S1: $g(n,c_1,c_{2^*},N)$}

In a closed linear chain, nodes can have values ``0'' or ``1''.
In this chain, a domain is a sequence of adjacent nodes with the same value (0 or 1), surrounded by nodes of different value.
In a chain of $N$ nodes, let $n$ be the total number of ones, $c_1$ the number of domains of ones preceded by a domain of only a zero and $c_{2^*}$ the number of domains of ones preceded by a domain of at least two zeros ( $c_{1}+c_{2^*}$ is the total number of domains of ones).
Recall Eq.~(\ref{eq:expected_S1_domains}), which gives the correlator of the protocol Subgraph S1 in terms of $n$, $c_1$ and $c_{2^*}$:
\begin{align*}
  \expected{\bm{K}_{\bm{x}}}
  &=
  (1-p_2)^{c_1+2c_{2^*}+n} (1-p_1)^{2(c_1+c_{2^*})}
  \notag\\
  &\quad
  \expected{K_{u_v}^{g_u}}^{c_1+c_{2^*}} \expected{K_u^{g_u}}^{n-c_1-c_{2^*}} \expected{K_u^{g_u}K_{u_v}^{g_u}}^{c_1+c_{2^*}}
  .
\end{align*}
Then, $g(n,c_1,c_{2^*},N)$ is the number of different configurations of that chain with given parameters, and
\[
  G(x,y_1,y_2,z) = \sum g(n,c_1,c_2,N) x^ny_1^{c_1}y_2^{c_{2^*}}z^N
\]
its generating function.
Each variable $x$, $y_1$, $y_2$, and $z$ ``counts'' the number of ones, domains of one zero, domains of two or more zeros and the total size, respectively.
The function $G$ can be found by joining simpler distributions.
We can think of the linear chain as a construction of domains of zeros and ones joined together.
Consider the set of domains of zeros, $\{0, 00, 000, \dots\}$, each domain of a given size appearing only once.
The number of domains of size $N$ in this set is $d_N=1$, which is generated by
\[
  D(z)
  =
  \sum_{n\ge1}d_Nz^N
  =
  \frac{z}{1-z}
  .
\]
We can differentiate between the set of domains of only a zero, $\{0\}$, and that of two or more zeros, $\{00, 000, \dots\}$.
In this case, the generating functions are respectively $z$ and $D(z)-z$.
The set of domains of ones, $\{1, 11, 111, \dots\}$, is generated by the same function, but here each element contributes to the total size of the chain and to the number of ones.
Its generating function is thus $D(xz)$.
The function generating the set of pairs of domains, the first of zeros and the second of ones, is
\[
  P
  \equiv
  P(x,y_1,y_2,z)
  =
  \left\{ y_1 z + y_2 \left[ D(z)-z \right] \right\} D(xz)
  .
\]
The function for all possible combinations of pairs of domains of zeros and ones (in order) is
\[
  \frac{1}{1-P}
  =
  1 + \left\{ y_1 z + y_2 \left[ D(z)-z \right] \right\} \frac{D(xz)}{1-P}
  .
\]
Here, the first element ($1$) counts the case where there is no pairs at all, and the second, $P/(1-P)$, to that where there is at least one pair.
We can now add nothing at all, a possible domain of zeros at the end, a possible domain of ones at the beginning, or both.
The final generating function is then
\begin{multline*}
  G
  =
  \left\{ 1 + \left[ y_1z + y_2\left( D(z)-z \right) \right] \frac{D(xz)}{1-P} \right\}
  \left[ 1 + D(xz) + P \right]
  \\
  +
  \left[ 1 + y_2D(z)D(xz)\frac{1}{1-P} \right] D(z).
\end{multline*}
Note that when we added a domain of zeros (the term with $D(z)$), we changed the variable $y_1$ for $y_2$, to take into account that the domain of zeros is now of size greater than one (because we are considering a closed linear chain).
Simplifying, we obtain
\begin{equation}
  G
  =
  \frac{1-xz^2(1-y_1+2(y_1-y_2)z)}{1-(1+x)z+x(1-y_1)z^2+x(y_1-y_2)z^3}
  .
\end{equation}
\subsection{Bipartite B: $g(n,c_1,c_2,c_{3^*},N)$}

In Bipartite B protocol, the  correlator depends on similar parameters, but here we have to differentiate between the number of domains of zeros with one element ($c_1$), two elements ($c_2$) and three or more elements ($c_{3^*}$):
\begin{align}
  \expected{\bm{K}_{\bm{x}}}
  &=
  (1-p_1)^n (1-p_2)^{c_1+2c_2+3c_{3^*}+n}
  \notag\\
  &\quad
  (1-2p)^n (1-p)^{2c_2+2c_{3^*}}
  .
\end{align}
This can be achieved by a small modification of the previous generating function.
Now, 
\begin{align*}
  P
  &\equiv
  P(x,y_1,y_2,y_3,z)
  \\
  &=
  \left\{ y_1 z + y_2 z^2 + y_3 \left[ D(z)-z-z^2 \right] \right\} D(xz)
  ,
\end{align*}
and
\begin{widetext}
\begin{equation}
  G
  =
  \frac{(1 + x z^2 (-1 + y_1 - 2 y_1 z + 2 y_2 z + 3 (-y_2 + y_3) z^2))}{(1 + z (-1 + x (-1 + z (1 + y_1 (-1 + z) + z (y_2 (-1 + z) - y_3 z)))))}
  .
\end{equation}
\end{widetext}

\subsection{Bipartite A: $g(n,c_1,c_{2^*},\bar{c}_{2^*},N)$}

In the Bipartite A, we also need to count the number $\bar{c}_{2^*}$ of domains of two or more ones.
The correlator in this case reads
\begin{align}
  \expected{\bm{K}_{\bm{x}}}
  &=
  (1-p_1)^{2n+2c_{2^*}+2\bar{c}_{2^*}}
  (1-p_2)^{2c_1+4c_2+2n}
  \notag\\
  &\quad
  (1-2p)^{2c_{2^*}+2\bar{c}_{2^*}} (1-p)^{n-2\bar{c}_{2^*}}
  .
\end{align}
We count $\bar{c}_{2^*}$ using variable $w_2$, and differentiating between the sets $\{1\}$ and $\{11,111,\dots\}$, which are generated by $xz$ and $D(xz)-xz$, respectively.
The extended function generating the set of pairs of domains is now
\begin{multline*}
  P
  \equiv
  P(x,y_1,y_2,w_2,z)
  =
  \left\{ y_1 z + y_2 \left[ D(z)-z \right] \right\}
  \\
  \left\{ x z + w_2 \left[ D(xz)-xz \right] \right\}
  .
\end{multline*}
Proceeding as in the previous case, we obtain
\begin{widetext}
\begin{equation}
  G
  =
  \frac{1+x z^2 (-1+y_1-2 y_1 z+2 y_2 z)+(-1+w_2) x^2 z^3 (y_1 (2-3 z)+3 y_2 z)}{1+z (-1+x (-1+z (1+(1+(-1+w_2) x z) (y_1 (-1+z)-y_2 z))))}
  .
\end{equation}
\end{widetext}

\subsection{Subgraphs S2: $g(n_{01},n_{10},n_{11},c_l,c_r,N)$}

In the Subgraphs S2 protocol, the sum is performed over even nodes.
In this case, it is convenient to express $\bm{x}$ as a sequence of elements 00, 01, 10 and 11 (the first digit corresponding to an odd node, and the second to an even node).
Each 01, 10 and 11 contribute to one \textsc{cphase} noise, as well as each domain of 00 which is preceded by a 01 or a 11.
Moreover, each 01 and 10 contribute to one $Y$ measurement noise.
Finally, each 01 and 11 contribute to a $1-3p$ noise of the purified subgraph, each 10 to a $1-p$ and each domain of 00 followed by a 10 or a 11 also to a $1-p$.
Thus, we need the number of configurations with $n_{01}$, $n_{10}$, and $n_{11}$ number of 01, 10 and 11 elements, and $c_l$ and $c_r$ domains of 00 preceded by 01 or 11 and followed by 10 or 11, respectively.
The correlator is
\begin{align}
  \expected{\bm{K}_{\bm{x}}}
  &=
  (1-p_2)^{n_{01}+n_{10}+n_{11}+c_l}
  (1-p_1)^{n_{01}+n_{10}}
  \notag\\&\quad
  (1-3p)^{n_{01}+n_{11}} (1-p)^{n_{10}+c_r}
  .
\end{align}

Now, each element contributes with $z^2$ to the size of the chain.
The domains $\{00,00\ 00,\dots\}$ are generated by $\frac{z^2}{1-z^2}$.
A domain made of elements 01, 10 and 11 of any size (including 0) is generated by
\[
  \frac{1}{1-(x_{01}+x_{10}+x_{11})z^2}
  ,
\]
and one which ends (or begins) with, say, element 01 (and thus is of size at least 2) is generated by
\[
  \frac{1}{1-(x_{01}+x_{10}+x_{11})z^2} x_{01}z^2
  .
\]
Proceeding as in the previous cases, the generating function is
\begin{equation}
  G
  =
  \left[ 1 + \frac{(x_{01}+x_{10}+x_{11})z^2}{1-(x_{01}+x_{10}+x_{11})z^2} + \frac{z^2}{1-z^2} + G \right] \frac{1}{1-P}
  ,
\end{equation}
where $P$ is
\begin{multline*}
  \left\{
  (x_{01}y_l + x_{10}y_r + x_{11}y_ly_r)z^2
  \right.
  \\
  \left.
  +
  \frac{1}{1-(x_{01}+x_{10}+x_{11})z^2} z^4 \left[ (x_{01}+x_{11})y_l+x_{10} \right]
  \right.
  \\
  \left.
  \left[ (x_{10}+x_{11})y_r+x_{01} \right]
  \right\}
  \frac{z^2}{1-z^2}
  .
\end{multline*}

\end{document}